\documentclass[ApJL,twocolumn]{aastex62}
\usepackage{amssymb, amsmath, graphicx, dcolumn, color,units, xspace, mathtools, physics, tensor, bm, lipsum, revsymb, booktabs, multirow}
\newcommand{\SPA}{School of Physics and Astronomy, Monash University, Clayton VIC 3800, Australia}
\newcommand{\OzGravMonash}{OzGrav: The ARC Centre of Excellence for Gravitational Wave Discovery, Clayton VIC 3800, Australia}
\newcommand{\LIGOlabMIT}{\affiliation{LIGO Laboratory, Massachusetts Institute of Technology, 185 Albany St, Cambridge, MA 02139, USA}}
\newcommand{\MKI}{\affiliation{Department of Physics and Kavli Institute for Astrophysics and Space Research, Massachusetts Institute of Technology, 77 Massachusetts Ave, Cambridge, MA 02139, USA}}

\shorttitle{White dwarf tides}
\shortauthors{Biscoveanu, Kremer, and Thrane}

\begin{document}

\title{Probing the efficiency of tidal synchronization in outspiralling double white dwarf binaries with LISA}

\author[0000-0001-7616-7366]{Sylvia Biscoveanu} \MKI \LIGOlabMIT

\author[0000-0002-4086-3180]{Kyle Kremer}
\affiliation{TAPIR, California Institute of Technology, Pasadena, CA 91125, USA}
\affiliation{The Observatories of the Carnegie Institution for Science, Pasadena, CA 91101, USA}

\author[0000-0002-4418-3895]{Eric Thrane}
\affiliation{\SPA}
\affiliation{\OzGravMonash}

\date{\today}

\begin{abstract}
{Compact-object binaries including a white dwarf component}
are unique among gravitational-wave sources because their evolution is governed not just by general relativity and tides, but also by mass transfer. 
While the black hole and neutron star binaries observed with ground-based gravitational-wave detectors are driven to inspiral due to the emission of gravitational radiation---manifesting as a ``chirp-like'' gravitational-wave signal---the astrophysical processes at work in double white dwarf (DWD) systems can cause the inspiral to stall and even reverse into an outspiral.
The dynamics of the DWD outspiral thus encode information about tides, which tell us about the behaviour of electron-degenerate matter.
We carry out a population study to determine the effect of the strength of tides on the distributions of the DWD binary parameters that the Laser Interferometer Space Antenna (LISA) will be able to constrain. 
{We find that the strength of tidal coupling parameterized via the tidal synchronization timescale at the onset of mass transfer affects the distribution of gravitational-wave frequencies and frequency derivatives for detectably mass-transferring DWD systems. Using a hierarchical Bayesian framework informed by binary population synthesis simulations, we demonstrate how this parameter can be inferred using LISA observations.} %
By measuring the population properties of DWDs, LISA will be able to probe the behavior of electron-degenerate matter.
\end{abstract}

\section{Introduction}
Double white dwarf (DWD) binaries detected with planned space-based gravitational wave detectors like the Laser Interferometer Space Antenna (LISA, \citealt{LISA:2017pwj}) offer a novel means with which to probe stellar physics and binary evolution processes (see \citealt{LISA2022} for a recent review). 
Such systems are the most common compact object binaries in the Milky Way~\citep{1995MNRAS.275..828M} and are expected to dominate LISA's confusion noise limit below $3\,$mHz~\citep{Nelemans:2000es, Nelemans:2001nr, Nelemans:2001hp, Ruiter:2007xx, Yu:2010fq, Liu:2014qaa}, with a predicted $\sim 10^{7}$ sources in the LISA band, $\sim 10^{5}$ of which are estimated to be resolvable~\citep{Ruiter:2007xx, Nissanke:2012eh, Lamberts:2019nyk}. Careful characterization of these systems is thus important not only to elucidate their population properties and evolution, but also to understand the sensitivity with which LISA will be able to detect other gravitational-wave sources, such as massive black hole binaries~\citep[e.g.,][]{Klein:2015hvg, Dayal:2018gwg, Bonetti:2018tpf, Barausse:2020mdt, Littenberg:2020bxy, Chen:2020qlp, Valiante:2020zhj}. 

As the white dwarfs in these systems are driven together via gravitational radiation, they may eventually get close enough to begin mass transfer. At this stage, the less massive of the two white dwarfs begins to overflow its Roche lobe and mass is pulled through the inner Lagrange point from the donor to the more massive accretor~\citep{1975MNRAS.172..493P, 1984ApJ...277..355W, 1984ApJS...54..335I, 2002ARep...46..667T}. 
If this mass transfer is dynamically unstable, a merger ensues which may ultimately lead to a Type Ia supernova for both systems with a total mass above the Chandrasekhar limit~\citep[e.g.,][]{1984ApJ...277..355W,1986ARA&A..24..205W, Maoz:2013hna} and sub-Chandrasekhar systems~\citep[e.g.,][]{Fink:2007fv, Fink:2010pz, Guillochon:2009hq, Kromer:2010iq, Shen:2017flp}. For systems with total mass near to or above the Chandrasekhar mass, the merger can also lead accretion-induced collapse of the remnant to a neutron star \citep[e.g.,][]{Miyaji1980,NomotoIben1985, Schwab:2015bma, Schwab2016}. 

On the other hand,
stable mass transfer---which in principle may proceed through either classic disk accretion~\citep[e.g.,][]{Soberman:1997mq, frank_king_raine_2002} or direct impact accretion~\citep[e.g.,][]{Marsh:2003rd}---can lead to the formation of an AM CVn system, a cataclysmic variable binary star characterized by accretion of hydrogen-poor material from a compact companion onto a white dwarf~\citep{1981ApJ...244..269N, Nelemans:2000es, 2002ARep...46..667T, 2010PASP..122.1133S}. The fate of interacting white dwarf systems remains highly uncertain due to poorly understood binary evolution processes, and it is possible that every mass transferring DWD system may eventually merge~\citep{2015ApJ...805L...6S}.

The stability of the mass transfer process depends sensitively on the properties of the system such as the white dwarf compositions, masses, orbital separation, accretor spin, and the strength of tidal coupling between the individual white dwarfs and the binary orbit~\citep{Marsh:2003rd, Gokhale:2006yn, Sepinsky:2014ila, 2015ApJ...806...76K}. Multi-messenger observations of stable systems, in the X-ray and optical bands and in gravitational waves, offer a unique opportunity to probe accretion physics. Such observations can also yield precise measurements of the binary orbital parameters, including their component masses~\citep{Nelemans:2003ha, 2014A&A...565A..11C, Breivik:2017jip, Korol:2018wep, Burdge:2020end, Johnson:2021grx}, due to their uniquely-determined evolutionary tracks~\citep{Deloye:2007uu, Kalomeni:2016hef}. Identifying galactic DWDs undergoing stable mass transfer with gravitational waves will facilitate their detection electromagnetically, since their intrinsic faintness has limited the number of currently-detected {AM CVn} systems in the galaxy to $\sim 65$, despite the fact that several thousand such sources are expected to exist in the Milky Way~\citep{vanRoestel:2022nqs, 2018A&A...620A.141R}.

The processes of mass transfer and tidal coupling change the evolution of the orbital frequency of stable systems, potentially leading to an increase in the orbital separation and a corresponding decrease in the frequency \citep[e.g.,][]{Kremer:2017xrg}, in contrast to the standard chirp-like increase in frequency due to gravitational-wave emission alone. Previous studies have estimated that up to several thousand such systems with a negative frequency derivative will be detectable with LISA, although the exact number depends on the specific binary evolution parameters used to model the population of DWDs in the Milky Way~\citep{Nelemans:2003ha, Ruiter:2007xx, Nissanke:2012eh, Kremer:2017xrg, Lamberts:2019nyk, Liu:2021tti} and the assumptions made when modeling the evolution of the system during mass transfer~\citep{Marsh:2003rd, Gokhale:2006yn, Sepinsky:2014ila, 2015ApJ...806...76K}. 

In this work, we use the DWD evolution prescription of \cite{Marsh:2003rd} in combination with the COSMIC population synthesis suite~\citep{Breivik:2019lmt} to determine the effect of the strength of tidal coupling on the properties of mass transferring DWD systems that will be detectable with LISA. The effect of this parameter has not been previously explored in works modeling both the evolution of mass transferring DWDs and a realistic Galactic population of these sources (e.g., \citealt{Kremer:2017xrg, Breivik:2019lmt,Seto:2022iuf}). We restrict our analysis specifically to systems with helium (He) white dwarf donors, which are expected to be the best candidates for producing stable mass-transferring systems. {We show that the effect of the strength of tidal coupling on the distribution of gravitational-wave frequencies and frequency derivatives, $(f_\text{GW}, \dot{f}_\text{GW})$, can be parameterized to {infer} the tidal coupling parameter with simulated LISA observations of the detectably mass-transferring population of DWDs.}
By measuring the shape of the $(f_\text{GW}, \dot{f}_\text{GW})$ distribution, space-based observatories like LISA will probe the physics of white dwarf tides, providing insights into the behavior of  electron-degenerate matter.

The remainder of this {manuscript} is organized as follows.
In Section~\ref{sec:methods}, we describe our methods for constructing a realistic population of mass-transferring DWDs in the Milky Way and their detection with LISA. The full details of the DWD evolution calculation are given in Appendix~\ref{ap:evolution}, and our evolution code is publicly available on \href{https://github.com/asb5468/dwd_mt_evolution}{git}. In Section~\ref{sec:results}, we present the results of our simulation for {six} different choices of binary evolution parameters and tidal coupling strength. {We demonstrate a hierarchical inference framework that can be used to directly constrain the strength of tidal coupling using LISA observations in Section~\ref{sec:inference}.} We conclude with a discussion of the results and a comparison with other works in Section~\ref{sec:conclusion}.

\section{DWD Population Synthesis and Evolution}
\label{sec:methods}
We use the COSMIC population synthesis suite~\citep{Breivik:2019lmt} to simulate a fiducial population of double white dwarf binaries. We use the default configuration settings except for the initial conditions, the convergence criteria, and the common envelope parameter $\lambda$, which we vary to account for the uncertainty in the common envelope physics. We draw sources from the \cite{Kroupa:2000iv} initial mass function, from a thermal eccentricity distribution~\citep{Heggie:1975tg}, and from a log-uniform separation distribution. Mass ratios are assumed to be distributed uniformly~\citep{1992ApJ...401..265M, 1994A&A...282..801G}, with the minimum mass ratio chosen such that the pre-main-sequence lifetime of the secondary star is not longer than the full lifetime of the primary if it were to evolve as a single star. We assume a binary fraction of 0.5.

The common envelope phase expected to precede the formation of the double white dwarf binary is a key theoretical uncertainty governing the properties of these systems at formation. The common envelope phenomenon occurs due to mass transfer from a giant star on a dynamical timescale, causing the Roche lobes of both the donor and its companion to overflow~\citep{Hurley:2002rf}. COSMIC adopts the standard ``$\alpha \lambda$'' common envelope formalism, where the parameter $\alpha$ determines the efficiency with which orbital energy is injected into the envelope, and $\lambda$ determines the binding energy of the envelope~\citep{Hurley:2002rf}. Our main results are presented for the choice of $\lambda=1$, but we also present results obtained with $\lambda=10$, to encapsulate some of the uncertainty due to common envelope physics~\citep{Dewi:2000nq}. We fix $\alpha=1$ in all simulations.

In order to ensure that the final converged population encompasses all the possible population synthesis outcomes for a fixed binary type, COSMIC compares histograms of the binary parameters between successive iterations of the simulation until a match criterion is met.
We use a match threshold of $10^{-4}$ for histograms of the donor and accretor masses, eccentricity, and orbital period between subsequent iterations, as described in \cite{Breivik:2019lmt, Kremer:2017xrg}. We evolve the initial simulated stellar population using COSMIC for 10 Gyr and compute the convergence of the sub-population comprising only those systems that have at least one He white dwarf component at the end of the evolution based on the final state. We limit the orbital periods of the converged population to the range $[2.3\times 10^{-5}, 0.23]$ days, so that they lie within the LISA sensitivity band: $f_\text{GW} \approx (\unit[0.1-1000]{mHz})$. From the converged population, we extract the donor and accretor masses at the first onset of Roche lobe overflow in the DWD systems with a He white dwarf donor, as systems with more massive donors are not expected to undergo stable mass transfer \citep[e.g.,][]{Marsh:2003rd}.

In order to more carefully resolve the timescales relevant for these white dwarf binaries as they begin mass transfer, we compute the subsequent binary evolution following the prescription of \citet{Marsh:2003rd}.
This amounts to solving a coupled system of first order differential equations encapsulating the effects of gravitational radiation (GR), mass transfer (MT), and tidal coupling for the component masses of the accretor and donor, $M_{A}$ and $M_{D}$, respectively, the accretor spin, $\Omega_{A}$, and the binary separation, $a$. Unlike the compact object binaries detectable with ground-based gravitational-wave detectors, the evolution of the binary orbit is not driven entirely by gravitational radiation. The total change in the orbital angular momentum can be expressed as the sum of the three aforementioned contributions:
\begin{align}
    \dot{J}_{\mathrm{orb}} = \dot{J}_{\mathrm{GR}} + \dot{J}_{\mathrm{MT}} + \dot{J}_{\mathrm{tides}}.
    \label{eq:ang_momentum}
\end{align}

In this work, we are specifically interested in determining the effect of tidal coupling between the white dwarf spin and the binary orbit on the stability of DWD mass transfer via its imprint on the distribution of the binary properties detectable with gravitational waves. We emphasize that unlike the tidal effects measured in the binary neutron star merger GW170817~\citep{2019PhRvX...9a1001A}, for example, the tides we consider here do not influence the orbital evolution of the binary by changing the phasing of the gravitational-wave signal (although see the discussion in Section~\ref{sec:conclusion}). Instead, the tides generate a dissipative torque term in Eq.~\ref{eq:ang_momentum} which drives the binary apart. 

We follow \cite{Marsh:2003rd} in parameterizing this effect via a synchronization timescale, which determines how long it takes for the spin of the white dwarf to become synchronized with the orbit due to tides.
The tidal strength, and hence the synchronization time, change as a function of the orbital separation and white dwarf masses~\citep{1984MNRAS.207..433C, Marsh:2003rd}, so we focus on exploring the effect of different values of the initial synchronization time at the onset of mass transfer, $\tau_0$. A smaller value of $\tau_0$ means stronger tidal effects. 

While considerable work has been done in modeling the effect of tides in DWDs~\citep{Fuller:2010em, Lai:2011xp, Fuller:2012dt, Fuller:2014ika, Willems:2009xk, Burkart:2012sq, Burkart:2013jua}, there remains significant uncertainty in the value of the initial synchronization timescale. We choose to explore the range between two extreme values---$10~\mathrm{yrs}$, corresponding to very efficient tidal coupling, and  $10^{15}~\mathrm{yr}$, corresponding to very weak coupling. {The binary evolution in the case of $\tau_0=10^{15}~\mathrm{yr}$ is expected to match the evolution when only the contributions from gravitational radiation and mass transfer are modeled, as the synchronizing torque is negligible for this choice of tidal timescale.} We further assume that the donor remains synchronized to the orbit throughout the evolution, but the accretor can be spun up due to accretion and resynchronized due to tides\footnote{See Section~\ref{sec:conclusion} for a discussion of the effect of this choice and other caveats for our tidal model.}. 

We evolve the systems for up to 10 Gyr from the onset of mass transfer but stop the evolution if $a < R_{A} + R_{D}$, {where $R_{A}$ and $R_{D}$ are the radii of the white dwarf accretor and donor, respectively}; if the mass transfer rate, $\dot{M}_{D}$, exceeds $0.01~M_{\odot}/\mathrm{yr}$; if the accretor mass exceeds the Chandrasekhar mass, $M_{\mathrm{Ch}}$; or if the donor mass drops below $0.01~M_{\odot}$. The first two scenarios will lead to a merger of the two white dwarfs, and the third scenario results in a Type Ia supernova \citep[e.g.,][]{1984ApJ...277..355W} or accretion induced collapse \citep[e.g.,][]{NomotoIben1985}. In the final scenario, as the donor approaches sufficiently low mass, Coulomb and thermal contributions to the donor's equation of state become important \citep{DeloyeBildsten2003} and our assumed white dwarf mass-radius relation is no longer appropriate. In practice, systems with such ultra-low-mass donors will likely not be detectable as LISA sources due to their relatively low chirp masses.

Our implementation differs from \cite{Marsh:2003rd} only in the treatment of super-Eddington accretion. While we do allow the accretor to spin asynchronously relative to the orbit, we use the approximation of \cite{1999A&A...349L..17H} to calculate the Eddington accretion rate, which assumes corotation of the accretor:
\begin{align}
\dot{M}_{\mathrm{Edd}} = \frac{8\pi c m_{p} G M_{A}}{\sigma_T(\phi_{L_1} - \phi_{A})},  
\end{align}
where $m_p$ is the proton mass, $\sigma_T$ is the Thomson cross section, and $\phi_{L_1} - \phi_{A}$ is the difference in the Roche potential at the inner Lagrangian point and the surface of the accretor. \cite{Marsh:2003rd} conclude that this approximation is sufficient immediately after the onset of mass transfer when the accretor is rotating slowly, which is the stage that has the biggest effect on the subsequent evolution of the acretion rate. We follow \cite{Kremer:2017xrg} and modify the evolution equations such that the accretion rate is capped at the Eddington limit,
\begin{align}
    \dot{M}_{A} = 
    \begin{cases}
    \dot{M}_{\mathrm{Edd}},\quad &\mathrm{if } \dot{M}_{D}\geq \dot{M}_{\mathrm{Edd}}\\
    \dot{M}_{D},\quad &\mathrm{otherwise}
    \end{cases},
\end{align}
instead of allowing for super-Eddington accretion and assuming that will lead to a merger. In this case, the mass transfer becomes non-conservative. We solve the coupled system of differential equations governing the orbital evolution using an 8th order Runge-Kutta integration scheme as implemented in \textsc{scipy}~\citep{2020SciPy-NMeth}. Further details on the evolution calculation are given in Appendix~\ref{ap:evolution}, and our evolution code is publicly available on \href{https://github.com/asb5468/dwd_mt_evolution}{git}. 

\begin{figure}
\centering
\includegraphics[width=0.45\textwidth]{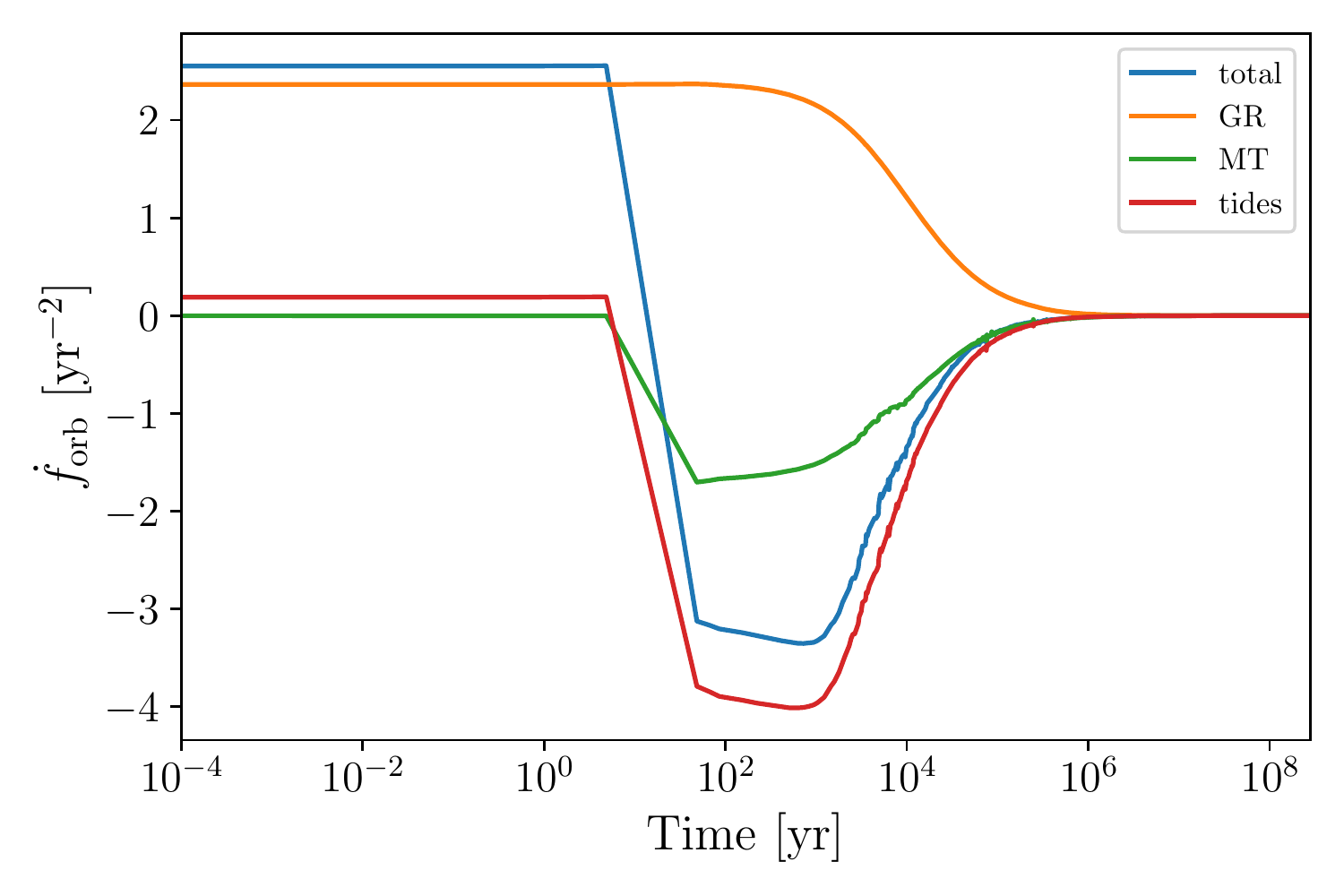}
\caption{Orbital frequency derivative for a system with $M_{D,0} = 0.25~M_{\odot},\ M_{A,0}=0.6~M_{\odot}$ that remains stable over 10 Gyr of evolution from the onset of mass transfer. While the contribution from gravitational radiation (GR, orange) is always positive, the terms stemming from mass transfer (MT, green) and tides (red) quickly become negative, such that the total frequency derivative (blue) also becomes negative, causing the orbital separation to increase as the system begins to outspiral. Observing such a system with LISA at a point in its evolution with $\dot{f}_{\mathrm{orb}}<0$ would provide insight into the effect of tides and the behavior of electron-degenerate matter. %
}
\label{fig:evolution}
\end{figure}

Figure~\ref{fig:evolution} shows the derivative of the orbital frequency as a function of time for a system with $M_{D,0} = 0.25~M_{\odot},\ M_{A,0}=0.6~M_{\odot}$ that remains stable for the entire 10 Gyr evolution time. The contribution of each of the three terms in Eq.~\ref{eq:ang_momentum} is shown individually. A smoothing function has been applied to suppress expected oscillations as described in Appendix~\ref{ap:oscillations}. {The GR contribution exhibits the characteristic ``chirping'' behavior where the emission of gravitational waves causes the orbital separation to shrink in order to conserve angular momentum, leading to a corresponding increase in the orbital frequency so $\dot{f}_{\mathrm{GR}} > 0$. However, the mass-radius relation for white dwarfs means that mass transfer from the less massive donor to the more massive accretor continuously drives the mass ratio to more extreme values, pushing the binary apart in order to conserve angular momentum so that $\dot{f}_{\mathrm{MT}} < 0$. The effect of tides is to redistribute the angular momentum of the accretor as it gets spun up back to the orbit, bringing the system to equilibrium and facilitating the stability of mass transfer. 

The total frequency derivative of the system,
\begin{align}
\dot{f}_{\mathrm{orb}} = \dot{f}_{\mathrm{GR}} + \dot{f}_{\mathrm{MT}} + \dot{f}_{\mathrm{tides}},
\end{align} 
remains positive until the onset of mass transfer causing the orbit to shrink, after which it becomes negative so the orbital separation actually increases and the frequency decreases as the binary ``outspirals.''\footnote{For our set-up (see Appendix~\ref{ap:evolution}) at fixed masses and accretor spin, the leading-order scaling of each of these three terms with the orbital separation, $a$, is $\dot{f}_{\mathrm{GR}} \propto a^{-11/2}, \dot{f}_{\mathrm{MT}} \propto a^0, \dot{f}_{\mathrm{tides}} \propto a^{-8}.$
}
Weaker tides are less efficient at redistributing the angular momentum back to the orbit, so the decrease in the total frequency derivative at the onset of mass transfer is less gradual, which can lead the accretion rate to become unstable in some cases.}

Figure~\ref{fig:evolution} highlights a key difference between DWDs as gravitational-wave sources and the neutron star and black hole binaries observed with ground-based gravitational-wave detectors. The frequency evolution of these interacting binaries is not dominated by gravitational radiation, but rather by the astrophysical effects of mass transfer and tides. {As we show in Section~\ref{sec:inference},} the observation of a population of systems with $\dot{f}_{\mathrm{orb}} <0$ with LISA can thus be used to constrain the synchronization timescale, $\tau_{0}$, which is related to the physics of electron-degenerate matter.

In order to map the converged COSMIC population onto a realistic Milky Way population of DWDs, we resample the fiducial population until we reach a thin disk stellar mass of $4.32 \times 10^{10}~M_{\odot}$~\citep{McMillan:2011wd}, simulating a uniform star formation rate by assigning a random birth time up to 10 Gyr for each zero-age main sequence (ZAMS) binary. We focus only on those systems that would be presently undergoing stable mass transfer for which $t_\mathrm{now} > \Delta t_\mathrm{birth} + \Delta t_\mathrm{DWD} + \Delta t_\mathrm{MT\ start}$, where $t_\mathrm{now} = 10~\mathrm{Gyr}$, $\Delta t_\mathrm{birth}$ is the ZAMS birth time of the binary, $\Delta t_\mathrm{DWD}$ is the delay time between birth and DWD formation, and $\Delta t_\mathrm{MT\ start}$ is the time between DWD formation and the first onset of mass transfer. We use COSMIC to track the evolution of each binary until $\Delta t_\mathrm{birth} + \Delta t_\mathrm{DWD} + \Delta t_\mathrm{MT\ start}$, at which point we continue to follow its evolution through the mass transfer phase using the \cite{Marsh:2003rd} prescription.

To reduce the computational cost of evolving each individual mass-transferring binary in the COSMIC population---as many have very similar masses---we grid the resulting $M_{A}, M_{D}$ mass distribution into bins of width $0.005~M_{\odot}$. Each of the binary bins is then evolved through the mass transfer phase following the prescription of \cite{Marsh:2003rd}. We then
use nearest-neighbor interpolation to assign each COSMIC binary to its corresponding mass bin. We remove binaries that merge due to unstable mass transfer, undergo a Type Ia supernova, or drop below the minimum white dwarf mass prior to $t_\mathrm{now}$ from the final population. 

For the systems remaining in the final population, we calculate the signal-to-noise ratio (SNR) with which they would be detected by LISA using the \textsc{legwork} package~\citep{Wagg:2021sgn}. We assume a 4-year LISA mission lifetime~\citep{LISA:2017pwj} and  corresponding sensitivity curve including confusion noise from galactic binaries in \cite{Robson:2018ifk}. 
We assign the position of each mass-transferring DWD throughout the thin disk by using COSMIC to sample from exponential profiles in both the radial and vertical directions with scale heights of $2.9~\mathrm{kpc}$ and $0.3~\mathrm{kpc}$, respectively~\citep{McMillan:2011wd}. 

We use the orbital frequency, $f_\mathrm{orb}$, when discussing the evolution of the white dwarf binary orbit, and the gravitational-wave emission frequency, $f_\mathrm{GW}$, when discussing gravitational-wave observations and detectability with LISA; the two are simply related via $f_{\mathrm{GW}} = 2f_\mathrm{orb}$. {We use the Fisher matrix formalism of \cite{Takahashi:2002ky} to calculate the approximate uncertainty in the measurement of $\dot{f}_{\mathrm{GW}}$ that would be obtained with LISA for each system and retain only those systems with detectably negative $\dot{f}_{\mathrm{GW}}$. This means the systems in our final population have $\mathrm{SNR}>7$,  $\dot{f}_{\mathrm{GW}}< 0~\mathrm{yr^{-2}}$, and $\Delta \dot{f}_{\mathrm{GW}}/|\dot{f}_{\mathrm{GW}}|\leq 0.5$, corresponding to a measurement that $\dot{f}_{\mathrm{GW}} < 0~\mathrm{yr^{-2}}$ at 95\% confidence.}

\section{Stability and Detectability with LISA}
\label{sec:results}
\begin{figure*}
	\centering
	\includegraphics[width=0.65\columnwidth]{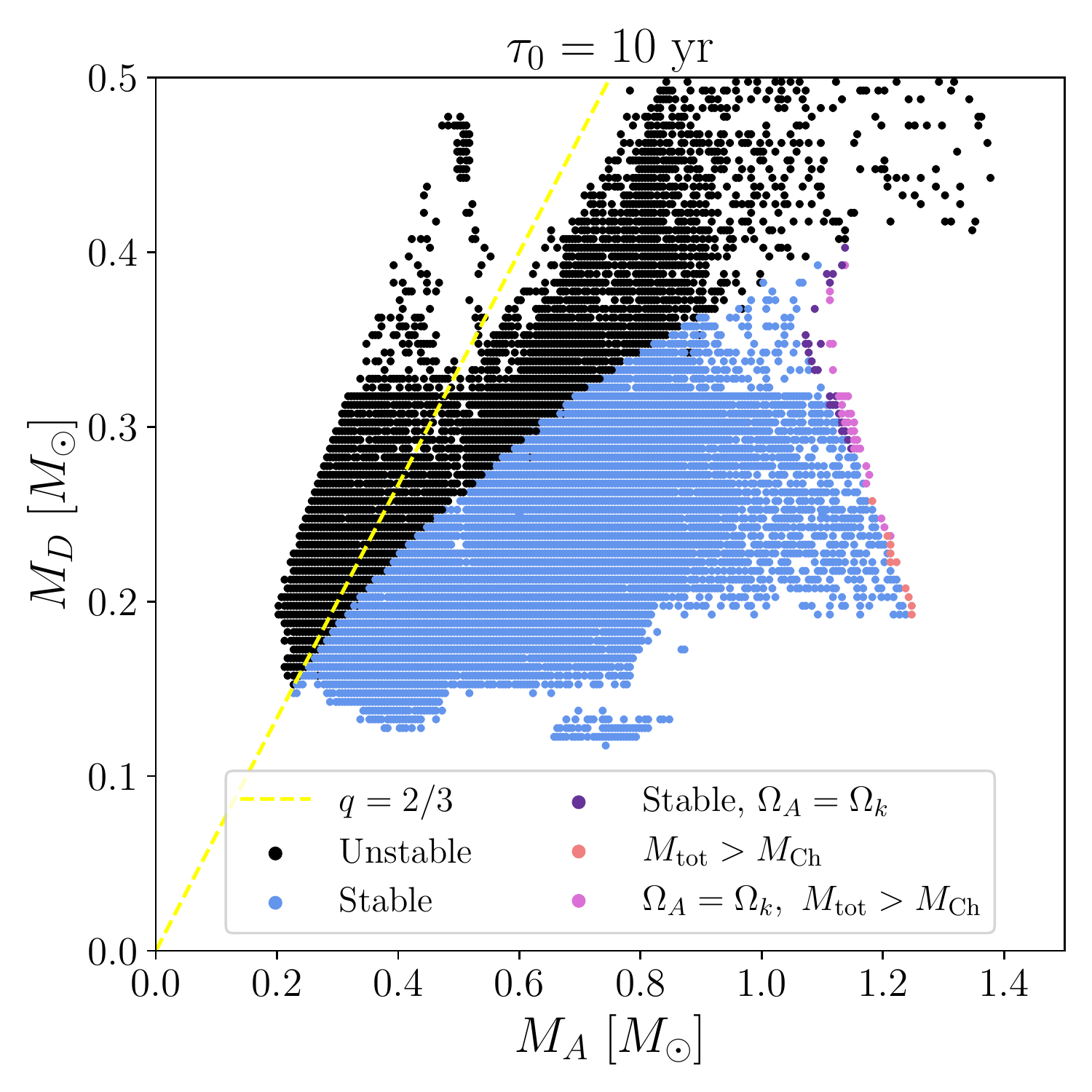}
	\includegraphics[width=0.65\columnwidth]{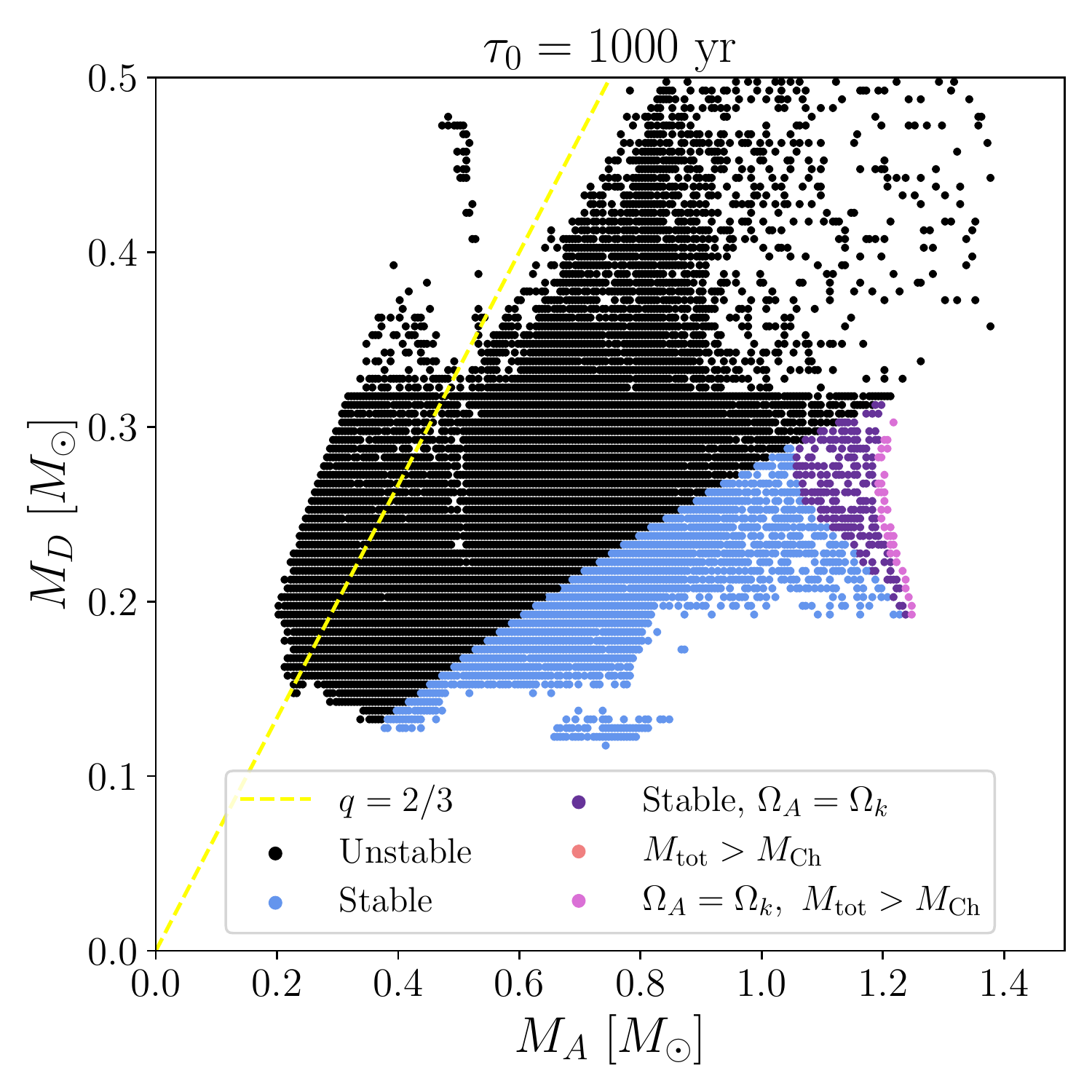}
	\includegraphics[width=0.65\columnwidth]{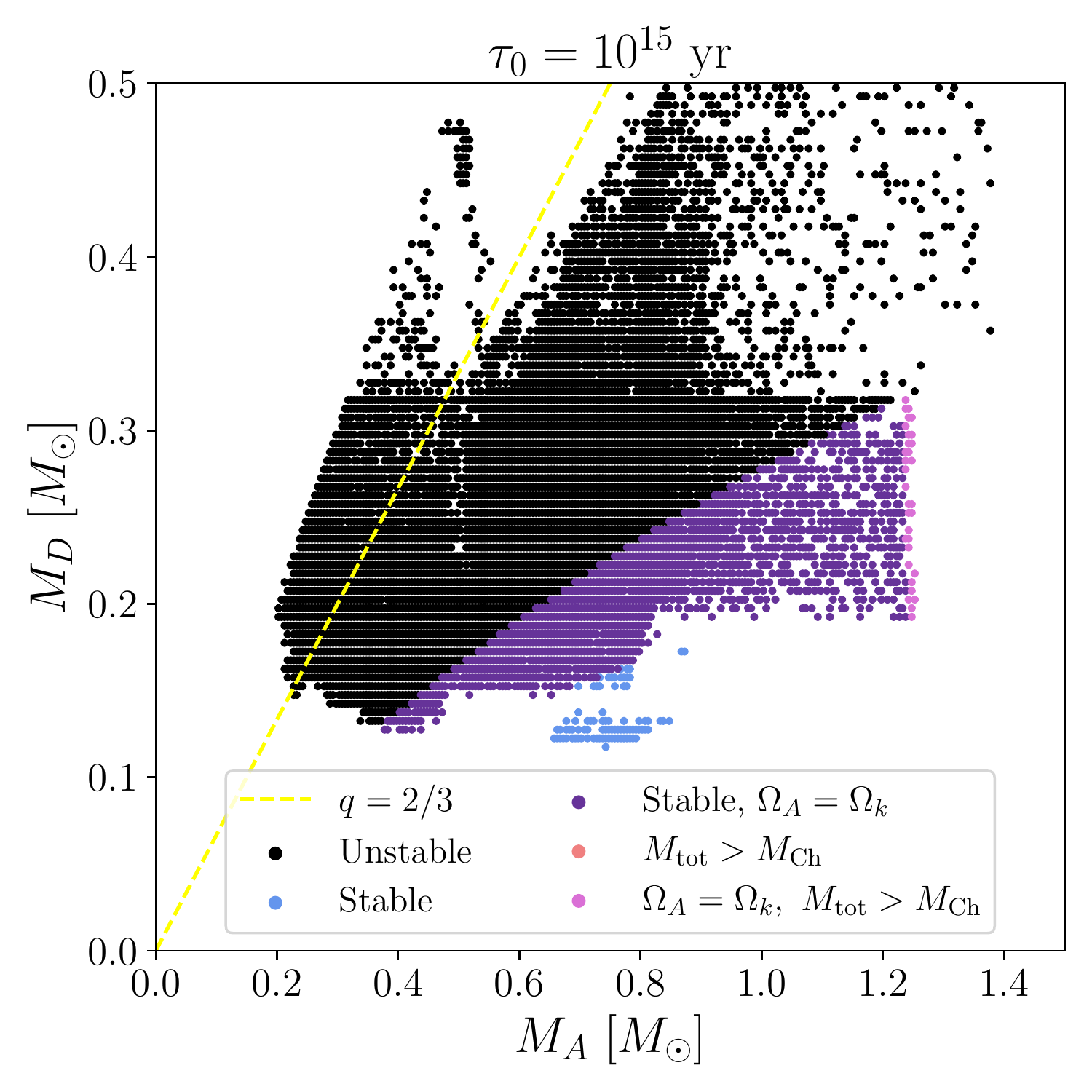}
	\caption{
	Scatter plot of the donor and accretor masses showing the stability of the systems corresponding to each of the mass bins in the combined base COSMIC populations for both $\lambda=1$ and $\lambda=10$ for $\tau_0 = 10~\mathrm{yr}$ (left), $\tau_0 = 1000~\mathrm{yr}$ (middle), and $\tau_0 = 10^{15}~\mathrm{yr}$ (right). {The mass bins corresponding to both values of $\lambda$ are plotted to provide more complete coverage of the parameter space.} Systems where the accretor reaches the breakup spin as described in Eqs.~\ref{eq:breakup1}-\ref{eq:breakup2} are denoted in the legend with $\Omega_{A}=\Omega_{k}$. Those where the total mass, $M_{\mathrm{tot}}=M_D + M_A$, exceeds the Chandrasekhar mass are indicated as $M_{\mathrm{tot}} > M_{\mathrm{Ch}}$. The canonical $q=2/3$ stability line is shown in yellow. Strong tides lead to stability over a larger region of the mass parameter space, and hence a larger number of mass-transferring systems detectable with LISA.}
	\label{fig:stability_grid}
\end{figure*}

\begin{figure*}
\centering
\includegraphics[width=0.8\textwidth]{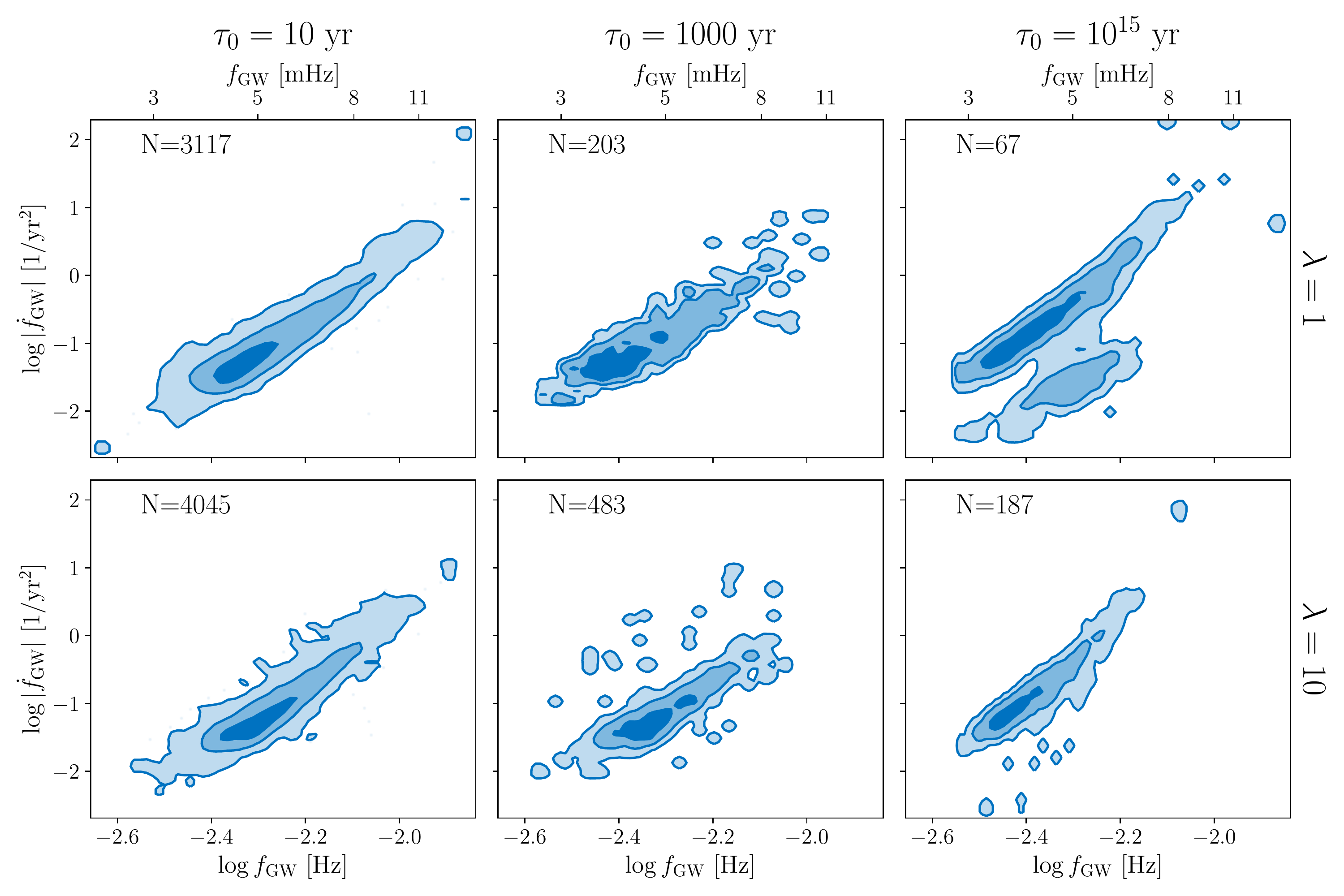}
\caption{Distribution of the log of the gravitational-wave frequency vs the log of the absolute value of the frequency derivative for systems with detectably negative $\dot{f}_{\mathrm{GW}}$ for different values of synchronization timescale, $\tau_0$ (columns), and common envelope parameter, $\lambda$ (rows). The shaded areas show the 1, 2, and $3\sigma$ credible regions, and the number of stable systems individually detectable with LISA is printed on each panel. The distribution of frequencies shifts to lower values as the synchronization time increases but does not change as significantly with $\lambda$. Comparison of the $(f_{\mathrm{GW}}, \dot{f}_{\mathrm{GW}})$ distribution of systems observed with LISA to these predictions can be used to constrain $\tau_{0}$ and $\lambda$.}
\label{fig:f_fdot}
\end{figure*}

In Fig.~\ref{fig:stability_grid}, we present a scatter plot of donor mass $M_D$ and accretor mass $M_A$, color-coded to indicate which systems undergo stable mass transfer, for the mass grid described in the previous section including systems from both the $\lambda=1$ and $\lambda=10$ populations. {The union of the grid points from the simulations with both choices of $\lambda$ is plotted to maximize the coverage of the mass parameter space.}
The left panel shows the results for strong tides ($\tau_0=10~\mathrm{yr}$), the middle panel shows the results for an intermediate synchronization timescale of $\tau_0=1000~\mathrm{yr}$, and the right panel shows the results for weak tides ($\tau_0=10^{15}~\mathrm{yr}$).
Systems with a total mass exceeding the Chandrasekhar mass that undergo a merger or Type Ia supernova immediately following the onset of mass transfer are omitted from the plot. Our results are consistent with those presented in \cite{Marsh:2003rd}. The shorter tidal synchronization time leads to a larger stable region of the parameter space, particularly for systems with more equal mass ratios. {This means that there are more stable systems with massive donors that come from more massive stellar progenitors in the case of stronger tides.} There are also significantly fewer systems where the accretor reaches the breakup spin, as strong tides are more efficient in resynchronizing the accretor spin with the orbit. 

The distributions of the logarithms\footnote{We use the base-10 logarithm throughout this manuscript.} of the gravitational-wave frequency and its derivative that would be observed by LISA for the final population obtained by convolving the stability grid shown in Fig.~\ref{fig:stability_grid} with the simulated population of DWD systems in the Milky Way thin disk as described in the previous section are shown in Fig.~\ref{fig:f_fdot}. Each column corresponds to a different value of the initial synchronization time, while each row corresponds to a different value of the $\lambda$ common envelope parameter. We only include systems with a detectably negative $\dot{f}_{\mathrm{GW}}$, reporting the number of such systems for each combination of $(\tau_0, \lambda)$ on the plot. 

As expected, the number of detectably interacting systems decreases with the tidal strength and falls off more steeply for the $\lambda=1$ population, as the mass distribution obtained with this choice of $\lambda$ covers a region of the parameter space with more unstable systems compared to $\lambda=10$. {The increase in the number of stable systems for the $\lambda=10$ simulations relative to the $\lambda=1$ case is due to two effects. First, the donor mass distribution at the onset of mass transfer for the base population with $\lambda=10$ peaks at lower masses, meaning more of these systems are likely to be stable. Second, increasing $\lambda$ decreases the binding energy of the envelope, meaning it is successfully ejected for more systems that can then go on to begin mass transfer as DWDs.} The number of systems with detectably negative $\dot{f}_{\mathrm{GW}}$ for the case of strong tides, $\tau_0=10~\mathrm{yr}$, is of the same order of magnitude as the results presented in \cite{Kremer:2017xrg}\footnote{The small variations in our results can potentially be explained due to differences in the binary evolution prescription implemented in COSMIC controlling the underlying DWD population at the onset of mass transfer and the treatment of the mass transfer physics, which is discussed in more detail in Section~\ref{sec:conclusion}.}.

While the choice of the common envelope parameter and synchronization timescale both affect the \textit{number} of LISA DWD systems with detectably negative $\dot{f}_{\mathrm{GW}}$, the shape of the $(f_{\mathrm{GW}}, \dot{f}_{\mathrm{GW}})$  distribution depends primarily on $\tau_0$, and does not change significantly with $\lambda$. The distribution of gravitational-wave frequencies shifts towards lower values with increasing tidal coupling strength. {To quantify this shift, we give the frequency and frequency derivative at the peak of 2D distribution for our six simulations in Table~\ref{tab:peaks}.} 

\begin{figure}
\centering
\includegraphics[width=0.5\textwidth]{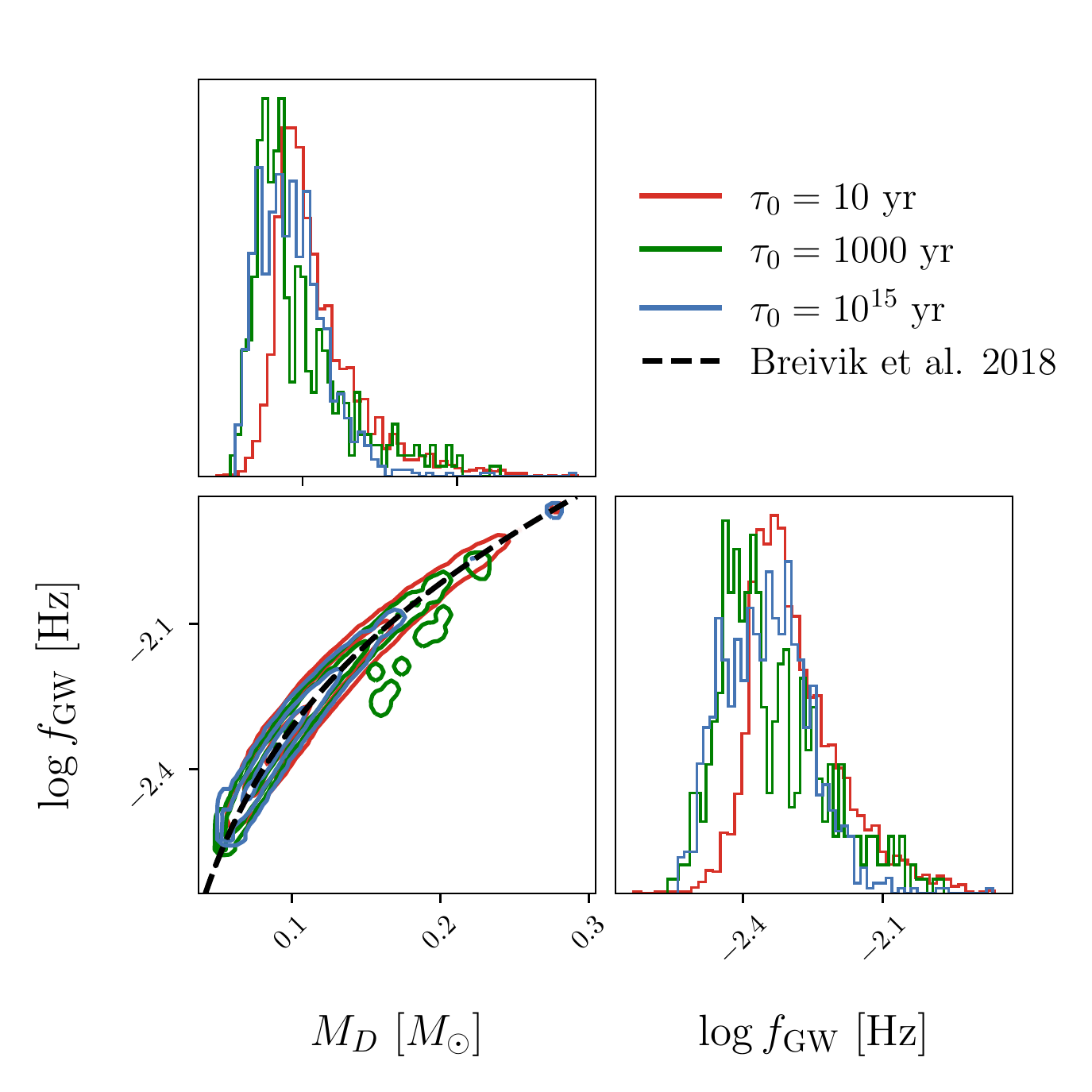}
\caption{Distribution of the donor mass vs the log of the gravitational-wave frequency for detectable systems with $\dot{f}_{\mathrm{GW}}\leq 1~\mathrm{bin/yr}$ for different values of synchronization timescale, $\tau_0$, and $\lambda=1$. The corner plot demonstrates the deterministic relationship between the donor mass and gravitational-wave frequency, which \cite{Breivik:2017jip} fit with a fourth-order polynomial, shown in the dotted black line. Decreasing the tidal synchronization time leads to more stable systems with higher donor masses and hence more systems with higher frequencies.
}
\label{fig:md_f}
\end{figure}

This effect is driven by the donor mass distribution of the systems undergoing stable mass transfer that are detectable with LISA due to the well-defined relationship between the donor mass and gravitational-wave frequency for these systems~\citep{Breivik:2017jip}, shown in Fig.~\ref{fig:md_f}. This relationship is a consequence of the relationship between orbital period and average density for semi-detached binaries undergoing Roche lobe overflow, $P \propto 1/\sqrt{\rho}$, since the density is purely a function of mass for white dwarfs with a known mass-radius relation. The width of the $(M_{D}, \log{f}_{\mathrm{GW}})$ distribution is due to a weak dependence of the period on the mass ratio of the system. Because stronger tides support the stability of systems with more massive donors, this deterministic relationship shifts the peak of the frequency distribution to higher frequencies. {While LISA observations are unable to \textit{directly} constrain the binary component masses and hence would not be able to produce the histogram of donor mass shown in Fig.~\ref{fig:md_f}, the tight relationship between frequency and donor mass means that if the frequency is well-measured, the donor mass can also be constrained for the detectably mass-transferring population.}
\begin{table}
    \centering
\begin{ruledtabular}
\begin{tabular}{l c c c c}
    $\tau_0$ [yr] & $\lambda$ & $\log f_{\mathrm{GW}}~[\mathrm{Hz}]$ & $f_{\mathrm{GW}}~[\mathrm{mHz}]$ & $\log |\dot{f}_{\mathrm{GW}}|~[1/\mathrm{yr}^{2}]$\\
    \midrule
    \multirow{2}{*}{10}& 1 & -2.34 & 4.57 & -1.39\\ & 10 & -2.30 & 5.01 & -1.35\\
    \multirow{2}{*}{1000}& 1 & -2.39 & 4.07 & -1.34 \\ & 10 & -2.32 & 4.79 & -1.27 \\
    \multirow{2}{*}{$10^{15}$}& 1 & -2.46 & 3.47 & -1.18 \\ & 10 & -2.44 & 3.63 & -1.15
\end{tabular}
\end{ruledtabular}
\caption{{Values of the log of the gravitational-wave frequency and frequency derivative at the peak of the 2D probability distribution for each of our six simulations.}}
\label{tab:peaks}
\end{table}

\section{Inference demonstration}
\label{sec:inference}
With LISA, a plot similar to Fig.~\ref{fig:f_fdot} can be constructed for the observed systems with detectably negative $\dot{f}_{\mathrm{GW}}$. By comparing the shape of the $(f_{\mathrm{GW}}, \dot{f}_{\mathrm{GW}})$ distribution and the number of detected systems to the predictions presented here, the values of $\lambda$ and $\tau_{0}$ can be constrained. Such a comparison can be performed using the framework of Bayesian hierarchical population modeling, where the hyper-parameters governing the $(f_{\mathrm{GW}}, \dot{f}_{\mathrm{GW}})$ distribution are inferred from observations {(see \cite{intro} for an introduction to hierarchical models in gravitational-wave astronomy). Here we demonstrate a method for performing this inference calculation.

The first step is to define a parameterization for the variation in the distribution of $(\log f_{\mathrm{GW}}, \log|\dot{f}_{\mathrm{GW}}|)$ as a function of $\tau_0$ in Fig.~\ref{fig:f_fdot}.\footnote{The units of the frequency and frequency derivative in this section are always Hz and $\mathrm{yr}^{-2}$, respectively, but we suppress the units going forward for ease of notation.}
For demonstration purposes, we choose a simple parameterization and model the correlation between $(\log f_{\mathrm{GW}}, \log|\dot{f}_{\mathrm{GW}}|)$ with a line, whose slope and intercept depend on the value of $\tau_0$ for each simulation.\footnote{
We do not attempt to simultaneously model the effects of both $\tau_{0}$ and $\lambda$ for this initial proof-of-concept, but emphasize that the effect of $\lambda$ and other physical parameters like temperature 
could be built into the model in an analogous way to the method we describe here for $\tau_0$ (see discussion in Section~\ref{sec:conclusion}).} 
After accounting for the correlation, we fit the degree of scatter about the line with a Gaussian distribution centered at $\mu=0$ and the distribution of the values of $\log|\dot{f}_{\mathrm{GW}}|$ predicted by the linear fit for each value of $\log f_{\mathrm{GW}}$ in the simulated population with a $\Gamma$ distribution. The free parameters of the model in this ``latent space'' are the slope and intercept of the line, the width of the Gaussian, and the shape, scale, and location of the Gamma distribution, $[a, b, \sigma, k, \theta, C] \in \Theta$, respectively (see Table~\ref{tab:individual_pars}).

The values of the parameters $\Theta$ for each population are uniquely determined by the value of the initial synchronization timescale, $\tau_0$. In order to infer the value of $\tau_0$ using an observed population, we need to create a mapping between $\Theta$ and $\tau_0$---in other words, a mapping from the latent space to the physical space. 
This mapping is the ``hierarchical model'', with ``hyper-parameters'' $\Lambda$, which are the same across all simulations no matter the value of $\tau_0$. The accuracy of the hierarchical model we develop is limited by the spacing of our grid of training simulations, which only includes three different values of $\tau_0$. 
Determining the optimal spacing for a simulation grid dense enough to robustly cover the parameter space is outside the scope of this work.
Moreover, we implicitly assume that the shape of the $(\log f_{\mathrm{GW}}, \log|\dot{f}_{\mathrm{GW}}|)$ does not depend strongly on the common envelope $\lambda$ parameter, and train our hierarchical model only on the $\lambda=1$ simulations. 
We thus emphasize that the results we obtain in this demonstration are not necessarily physically realistic in terms of constraining power and are merely meant to illustrate the method by which such inference can be done in the future with a more detailed training grid.

\begin{figure*}
\centering
\includegraphics[width=0.32\textwidth]{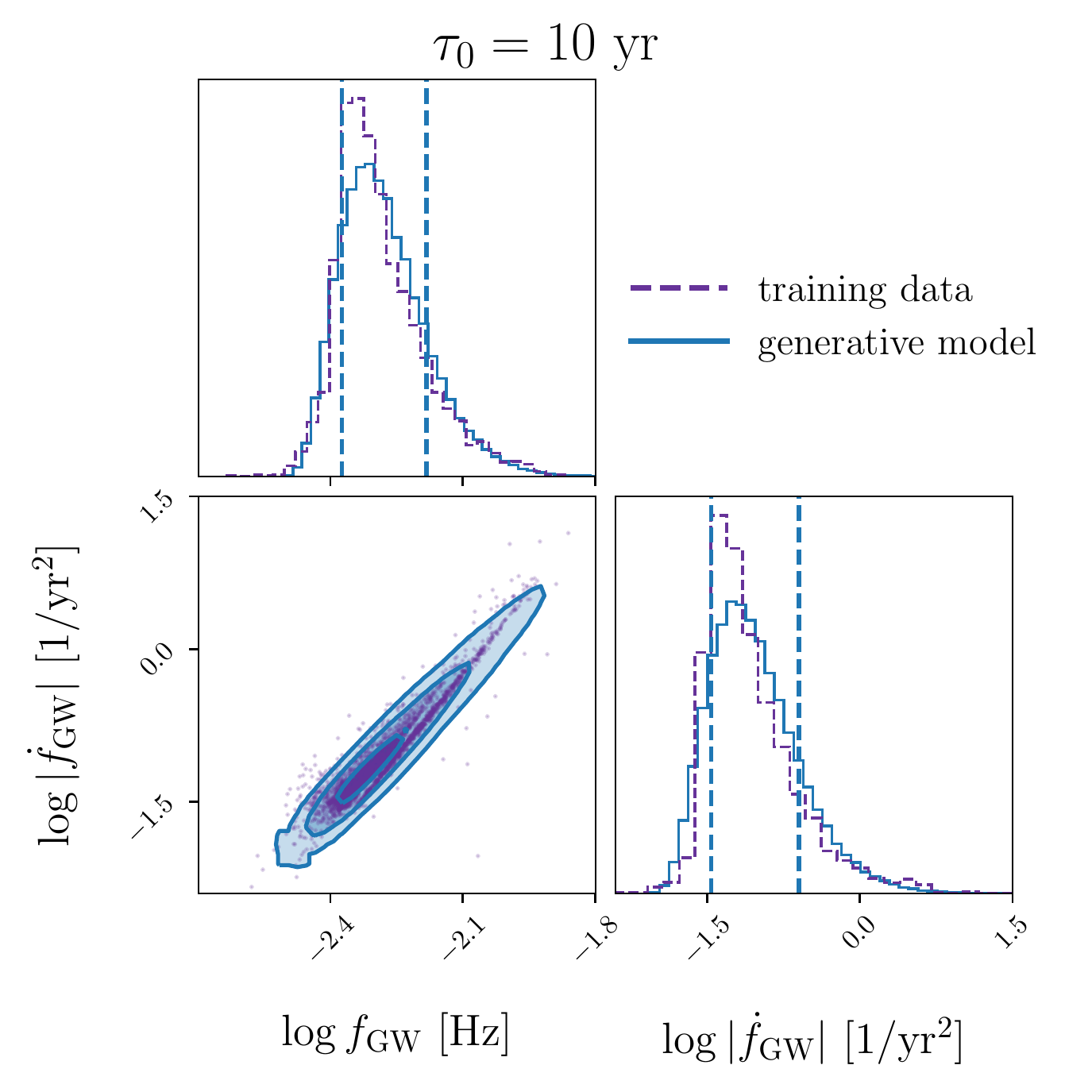}
\includegraphics[width=0.32\textwidth]{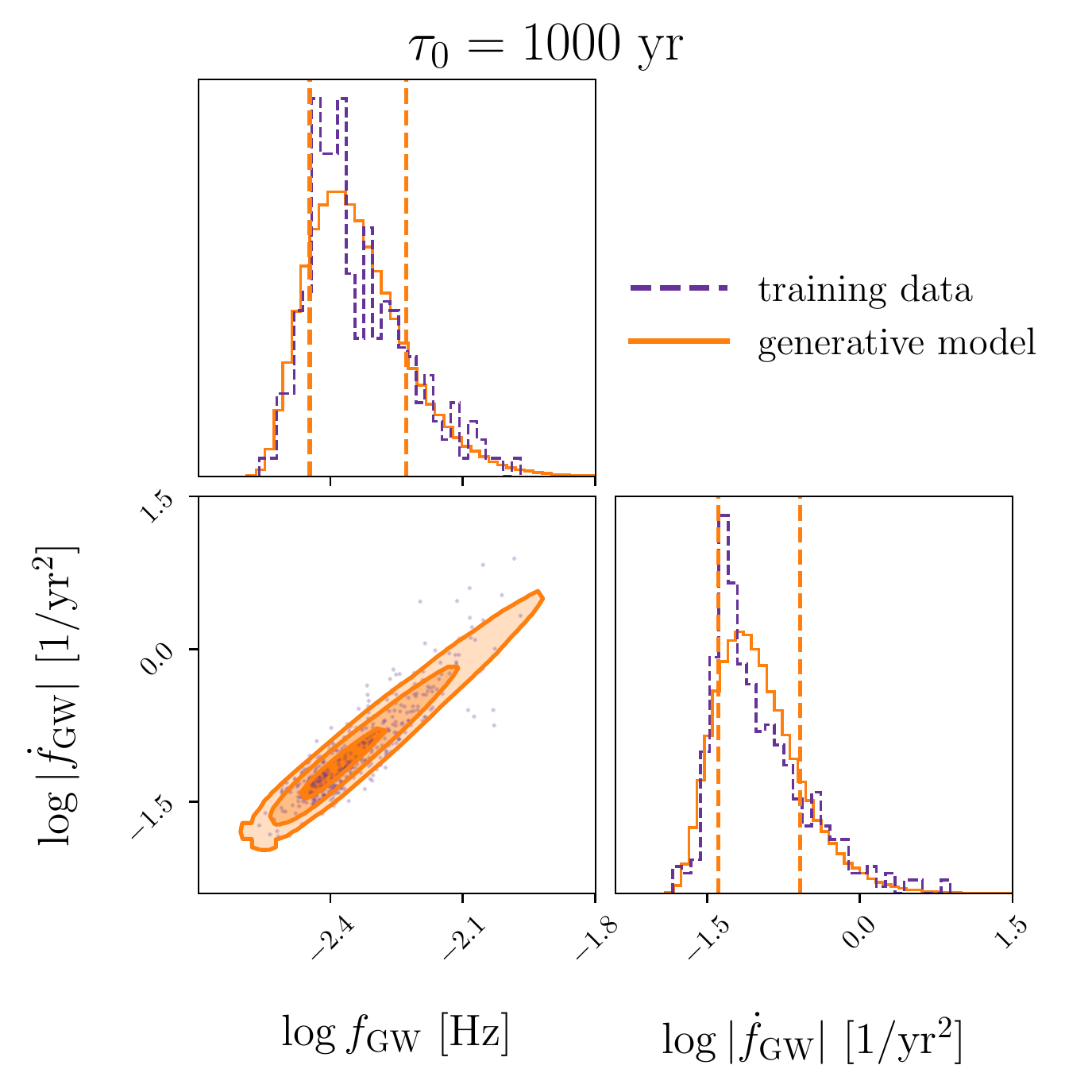}
\includegraphics[width=0.32\textwidth]{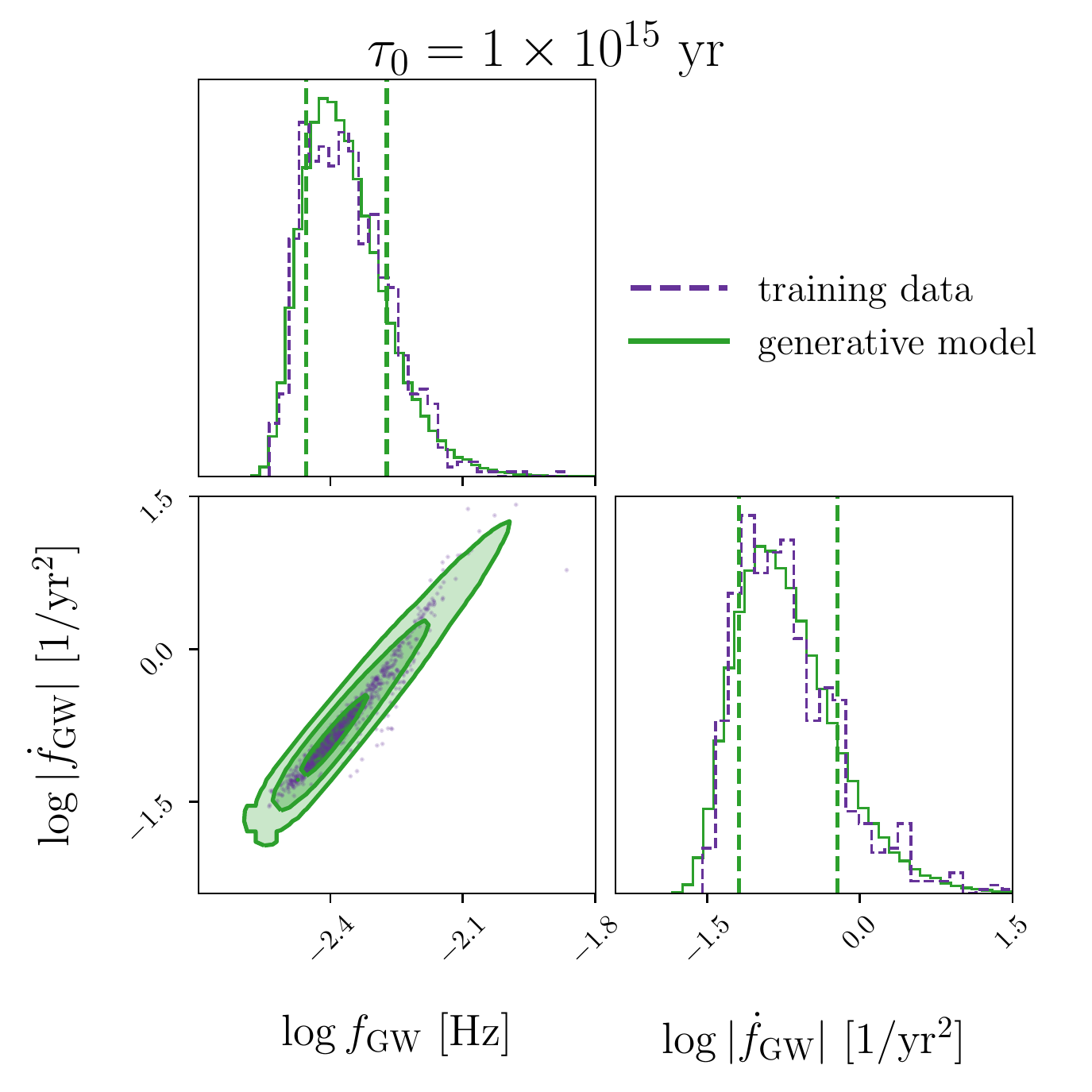}
\caption{{Comparison distributions of $(\log f_{\mathrm{GW}}, \log | \dot{f}_{\mathrm{GW}}|)$ obtained using the generative model marginalized over parameter uncertainty and the original simulation used for training for the three different values of $\tau_0$ considered for the $\lambda=1$ population. The model assumes the correlation between $\log f_{\mathrm{GW}}$ and $\log | \dot{f}_{\mathrm{GW}}|$ is linear, that the distribution of perpendicular distances of the simulation points from the line is Gaussian, and that the value of $\log | \dot{f}_{\mathrm{GW}}|$ predicted by the linear fit for each value of  $\log f_{\mathrm{GW}}$ follows a Gamma distribution. The side lobe of samples in $(\log f_{\mathrm{GW}}, \log |\dot{f}_{\mathrm{GW}}|)$ for the $\tau_0=10^{15}~\mathrm{yr}$ simulation has been removed from the training data, as this region is likely artificially overpopulated by our base-population resampling procedure.}}
\label{fig:generative_comp}
\end{figure*}

We use hierarchical Bayesian inference to fit the hyper-parameters $\Lambda$ of our generative model. The full details of the calculation, including the likelihood, prior, and sampler choices, are given in Appendix~\ref{ap:inference}.
The values of $\Theta$ implied by our inferred values of $\Lambda$ for each of our three training simulations are provided in Table~\ref{tab:individual_pars}.
The inferred $\Theta$ values have statistically significant differences for each value of $\tau_0$, with the uncertainty coming from the uncertainty in the $\Lambda$ posterior.  
Figure~\ref{fig:generative_comp} shows the comparison of the three simulated distributions with the prediction from the generative model described above marginalized over the uncertainty in the inferred values of $\Theta$ for each simulation. {To quantify the degree of statistical similarity between the distributions predicted by the generative model and the training data, we compute the Jensen-Shannon (JS) divergence~\citep{js_divergence} between the one-dimensional frequency and frequency derivative distributions for each choice of $\tau_0$. The JS divergence ranges from 0 to 0.69 nats, the natural unit of information, for identical distributions to maximally divergent distributions. 
Previous applications of this statistic in gravitational-wave astronomy conclude that values below 0.002 nats generated with thousands of samples correspond to statistically consistent distributions~\citep{Romero-Shaw:2020owr}. We find values in the range $[0.009, 0.002]~\mathrm{nats}$, indicative of some statistical differences between the generative model and the training data. Nonetheless, we believe the generative model captures the salient features required for our illustrative purposes.}

\begin{table*}
\begin{ruledtabular}
\begin{tabular}{l l l l l}
    Parameter & Description & $\tau_0=10~\mathrm{yr}$ & $\tau_0=1000~\mathrm{yr}$ & $\tau_0=10^{15}~\mathrm{yr}$\\
    \midrule
    $a$ & Slope of linear correlation & $4.50^{+0.09}_{-0.09}$& $3.67^{+0.13}_{-0.11}$ & $5.28^{+0.11}_{-0.11}$ \\
    $b$ & Intercept of linear correlation & $9.22^{+0.21}_{-0.21}$& $7.60^{+0.30}_{-0.26}$& $11.78^{+0.26}_{-0.27}$\\
    $\sigma$ & Width of Gaussian distribution about line & $28.3^{+1.7}_{-1.6}\times 10^{-3}$& $56.1^{+3.5}_{-3.2}\times 10^{-3}$& $25.6^{+1.6}_{-1.4}\times 10^{-3}$ \\
    $k$ & $\Gamma$-distribution shape & $5.82^{+1.6}_{-1.0}$ & $4.9^{+1.1}_{-0.7}$ & $5.4^{+1.5}_{-1.7}$\\
    $\theta$ & $\Gamma$-distribution scale & $18.5^{+2.2}_{-2.4}\times 10^{-2}$ & $18.9^{+2.1}_{-2.2}\times 10^{-2}$& $21.5^{+3.7}_{-2.8}\times 10^{-2}$\\
    $C$  & $\Gamma$-distribution offset & $-2.10^{+0.10}_{-0.14}$& $-1.92^{+0.07}_{-0.10}$& $-1.86^{+0.11}_{-0.15}$ \\
\end{tabular}
\end{ruledtabular}
\caption{{Parameter descriptions, inferred medians, and symmetric 90\% credible intervals for the model characterizing the distribution of $(\log f_{\mathrm{GW}}~[\mathrm{Hz}], \log|\dot{f}_{\mathrm{GW}}|~[1/\mathrm{yr^{2}}])$ for each value of $\tau_0$ in our simulations with $\lambda=1$.}}
\label{tab:individual_pars}
\end{table*} 

\begin{figure}
\includegraphics[width=\columnwidth]{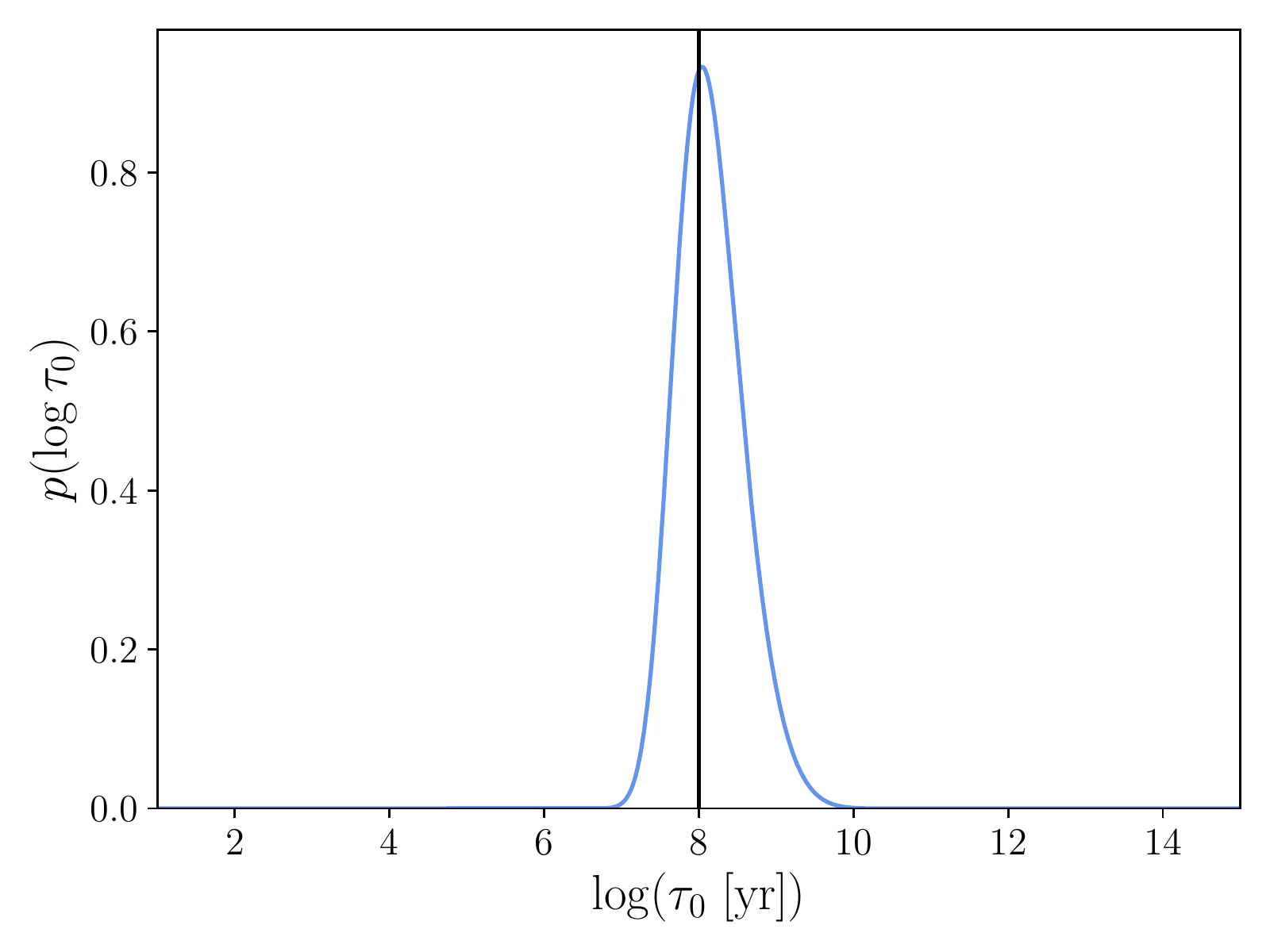}
\caption{{Posterior on the log of the initial synchronization timescale for a simulated population of 100 interacting DWDs detectable with LISA. The black vertical line indicates the true value of $\log\tau_{0}$ used to simulate the population.}}
\label{fig:tau_post}
\end{figure}

\begin{figure*}
\includegraphics[width=\textwidth]{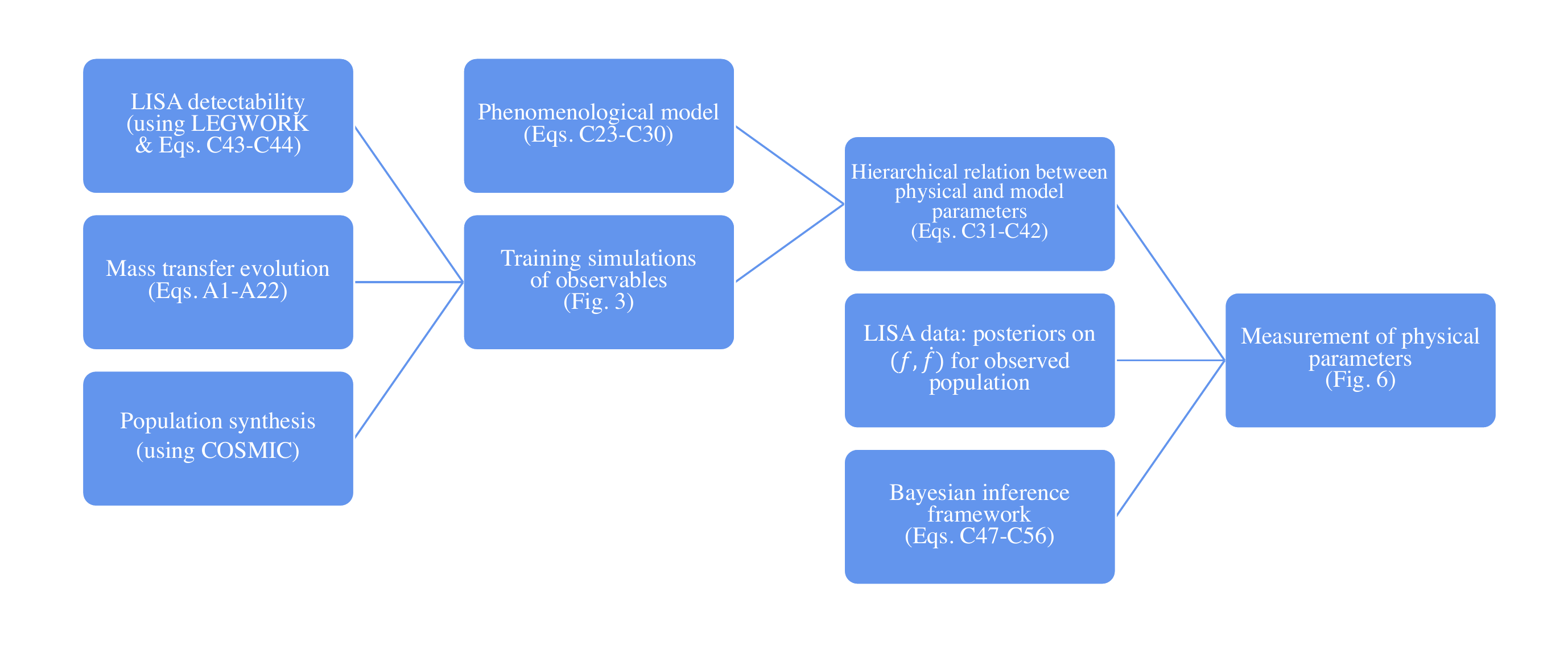}
\caption{Flow chart demonstrating the process for combining LISA data (real or simulated, as in this work) with our population synthesis simulations to arrive at the constraint on the initial tidal synchronization timescale presented in Fig.~\ref{fig:tau_post}.}
\label{fig:flow_chart}
\end{figure*}

With our generative model in hand, we now demonstrate how $\tau_0$ can be inferred using a simulated population based on this model with $\tau_{0} = 10^{8}~\mathrm{yr}$. 
We generate simulated LISA measurements of $(\log f_{\mathrm{GW}}, \log|\dot{f}_{\mathrm{GW}}|)$, modeling the statistical uncertainty with the Fisher matrix formalism derived in \cite{Takahashi:2002ky}. 
We then combine the individual measurements of $(\log f_{\mathrm{GW}}, \log|\dot{f}_{\mathrm{GW}}|)$ to estimate $\tau_{0}$ hierarchically.\footnote{The resulting posterior distributions for $\tau_0$ include not only measurement uncertainty, but also theoretical uncertainty from our generative model fit.} More details on this second inference step are presented in Appendix~\ref{ap:inference}.
Figure~\ref{fig:tau_post} shows the posterior on $\log\tau_{0}$ that we obtain for a simulated population of 100 interacting double white dwarfs detectable with LISA using the same selection criteria applied to the final populations in our population synthesis models. For a population of this size, the width of the 90\% credible interval on $\log_{10}\tau_{0}$ is 1.44 dex. For a lower value of the initial synchronization timescale, the number of detectable systems would be higher, and hence the uncertainty on $\log_{10}\tau_{0}$ would be smaller.
Since we started with uniform prior on $\log\tau_0$ over the range spanned by the x-axis of Fig.~\ref{fig:tau_post}, this measurement represents a significant improvement in our knowledge of the tidal timescale.} {The flow chart in Fig.~\ref{fig:flow_chart} demonstrates the process for combining our population synthesis simulations presented in Fig.~\ref{fig:f_fdot} with the hierarchical framework described in this Section and in Appendix~\ref{ap:inference} to produce the posterior on the initial tidal synchronization timescale shown in Fig.~\ref{fig:tau_post}.}

\section{Discussion}
\label{sec:conclusion}

Unlike the neutron star and black hole binaries observed by ground-based gravitational-wave detectors so far, the orbital evolution of double white dwarf systems observable with space-based detectors like LISA can be strongly affected by phenomena other than gravitational waves. For interacting DWDs, mass transfer and tides also contribute to the change in the orbital angular momentum. Because the radius of the less massive white dwarf donor grows as it loses mass, the binary separation increases, causing the the system to ``outspiral'' with a negative orbital and gravitational-wave frequency derivative. The strength of the tidal coupling between the accretor spin and the binary orbit determines which regions of the donor and accretor mass parameter space support stable mass transfer. Thus, the dynamics of the outspiral encode information about the tides, and by measuring the population properties of DWDs with LISA, we can probe the physics of electron-degenerate matter.

In this work, we have demonstrated that the distribution of gravitational-wave frequencies and frequency derivatives, $(f_{\mathrm{GW}}, \dot{f}_{\mathrm{GW}})$, for DWD systems detectable with LISA 
can be used to constrain the strength of tidal coupling between the accretor spin and the orbit.\footnote{We have implicitly assumed that mass transfer is the only process that could drive a DWD system to outspiral. It is possible that other mechanisms such as environmental effects~\citep{Barausse:2014pra, Cardoso:2019rou}---which have yet to be studied in the context of DWDs---could lead to a similar reversal of frequency derivative of the binary. Such effects may in principle interfere with the ability to constrain DWD tidal physics, although they are highly unlikely in the population of Milky Way binaries accessible to LISA.}
The shape of this distribution is determined by the initial synchronization timescale $\tau_0$, while the total number of interacting DWD systems can be used to inform common envelope physics via the parameter $\lambda$.

Future work should seek to incorporate a more complete model of the tidal physics governing the stability of mass transfer in interacting DWDs in order to obtain the most accurate estimates of the tidal hyper-parameters from the population observable with LISA. While our model is adequate for showing qualitatively how LISA will be able to measure $(\tau_{0}, \lambda)$, a few simplifying assumptions could be relaxed to achieve a more physically realistic model.

First, we assume that the donor spin remains synchronized to the orbit throughout the entire process of stable mass transfer, ignoring the tidal torque term that arises due to the asynchronicity of the donor. \cite{Gokhale:2006yn} take this term into account but find only a minor increase in the region of the mass parameter space that supports stable mass transfer. We also ignore the effect of the asynchronicity of both the donor and accretor on the geometry of the Roche lobe, which was found in \cite{2015ApJ...806...76K, Sepinsky:2007in} to marginally increase the number of stable systems for the case of strong tides only. 

Finally, we rely on the approximation of \cite{1988ApJ...332..193V}, which stipulates that the angular momentum transferred from the orbit to the accretor is exactly equal to the angular momentum of a ballistic particle at its average radius along its trajectory from the donor to the accretor during mass transfer. By implementing the more detailed calculations of \cite{Sepinsky:2007in} tracking the ballistic trajectory of each transferred mass particle, \cite{Sepinsky:2014ila, 2015ApJ...806...76K} find a further increase in the region of the mass parameter space supporting stable mass transfer. The effects of each of these assumptions is summarized in Figs. 5-6 of \cite{Kremer:2017xrg}. While relaxing these various assumptions would lead to an increase in the number of stable systems in our final population, this does not affect the main qualitative conclusion of our analysis---that the tidal synchronization timescale leaves a clear imprint on the detectable distribution of $(f_{\mathrm{GW}}, \dot{f}_{\mathrm{GW}})$ for LISA.

{We have also neglected the effect of temperature on the white dwarf mass-radius relation, with higher temperatures generally leading to larger radii~\citep{2000A&A...353..970P}. Studies conducting full evolution of low-mass white dwarfs find that the radius can be a factor of four larger for astrophysically realistic temperatures~\citep{2007MNRAS.382..779P}. This would translate into roughly a factor of eight decrease in the gravitational-wave frequency, an effect that could be confused with increasing the tidal synchronization timescale, as shown in Fig.~\ref{fig:f_fdot}. However, the distribution of temperature in the population of DWD systems observable with LISA can be built into our model as an additional physical parameter, like the $\lambda$ common envelope parameter. This would require running a larger suite of simulations including the effect of temperature and adjusting the generative model to account for the effect of this parameter on the latent space variables. The functional form of the mapping between the latent space and the physical space could be inferred using symbolic regression or normalizing flows~\citep[e.g.,][]{Wong:2022bxp, Ruhe:2022ddi} to accommodate more degrees of freedom. We emphasize that, while introducing a distribution of temperatures might broaden the features of the $(\log f_{\mathrm{GW}}, \log | \dot{f}_{\mathrm{GW}}|)$ distributions shown in Fig.~\ref{fig:f_fdot}, the effect of changing the tidal synchronization timescale should still be distinguishable in the population.}

Parameterizing all tidal effects at work during DWD evolution via a single synchronization timescale obfuscates the contributions of the complicated individual tidal processes. Tidal effects can broadly be broken down into two categories---equilibrium tides, where a static bulge is induced in the star, and dynamical tides, which consist of stellar oscillations. Dynamical tides can either be linear, operating as discrete standing waves (``g-modes'')~\citep{Fuller:2010em}, or non-linear, manifesting as traveling waves that propagate towards the surface of the white dwarf where they are likely dissipated~\citep{Lai:2011xp}. For the DWD binaries in our final population with orbital periods in the range of $3 \lesssim P_{\mathrm{orb}}/\min \lesssim 13$, we expect the quasi-static~\citep{Willems:2009xk}, linear~\citep{Fuller:2010em, Burkart:2012sq, Burkart:2013jua} and weakly non-linear~\citep{Yu:2020bic} approximations to break down such that the tidal effects are dominated by strongly nonlinear processes. 

Previous work has shown that non-linear, dynamical tides can play a significant role in the synchronization of the white dwarf spin with the binary orbit and can affect the frequency evolution of the gravitational-wave signal accessible with LISA~\citep{Lai:2011xp, Fuller:2012dt, Fuller:2014ika, Yu:2020bic}. \cite{McNeill:2019rct} {find that non-resonant dynamical tides} %
can induce eccentricity in the system and modify the gravitational waveform. Tidal effects are also found to lead to significant tidal heating~\citep{Piro:2011qe, Fuller:2012ky, Fuller:2012dt} that may affect the binary orbit, or otherwise increase the tidal coupling strength via the dependence of the model we adopt here based on \cite{1984MNRAS.207..433C} on the white dwarf luminosity. 

Because of the significant uncertainties in these tidal models used to directly estimate the tidal timescale, the approach we employ in this work should be taken as a first step towards modeling the effects of tides on the observable population properties of interacting DWDs by leaving the synchronization timescale as a free parameter. Development of an improved physical parameterization of the dissipative tidal torque in Eq.~\ref{eq:ang_momentum} could lead to direct constraints on these specific non-linear, dynamical tidal effects rather than the generalized synchronization timescale via hierarchical inference of the LISA interacting DWD population in the future. Furthermore, we have assumed $\tau_0$ to be constant and universal across the entire white dwarf population. However, its dependence on stellar structure~\citep{1984MNRAS.207..433C} implies that it may vary over the course of a binary's evolution due to the effects of mass loss or tidal heating on the white dwarf density profile. It should also vary across the distribution of temperatures in the white dwarfs population. Incorporating these more subtle physical effects into future applications of the framework we have illustrated here will lead to improved constraints on the tidal processes at play in white dwarfs. 

Some models of DWD evolution alternatively predict that the initial phase of mass transfer of the non-degenerate outer layer of hydrogen in low-mass donor white dwarfs may lead to classical nova-like outbursts~\citep{DAntona:2006wor, Kaplan:2012hu, 2013ApJ...770L..35S}. Dynamical friction in the expanding nova shells shrinks the orbit, leading to an increase in the accretion rate. This behavior continues through the subsequent accretion of the degenerate He layers, causing the accretion to become unstable and leading to the eventual merger of the system~\citep{2015ApJ...805L...6S}. These models predict that every interacting double white dwarf eventually merges, implying that none of the systems we consider in this work would be stably accreting and observable with LISA. Thus, the detection and characterization of such systems with LISA will help to discriminate between the stable accretion and rapid merger models. %

Complementary electromagnetic measurements of the properties of individual AM CVns will also allow us to constrain the processes affecting the orbital evolution of these systems. The three shortest-period known interacting DWD systems (which will all be observable with LISA) all have measurements of their period derivative, obtained via detailed photometric observations and timing of their light-curves. The system ES Ceti is observed to be outspiraling, with $\dot{P}=3.2\times 10^{-12}$, consistent with the prediction for a DWD system undergoing disk accretion~\citep{2018ApJ...852...19D}, showing that at least some interacting DWDs do indeed avoid merger.

The other two systems, HM Cancri (HM Cnc, also known as RX J0806.3+1527) and V407 Vul (RX J1914.4+2456), however, have positive $\dot{f}_{\mathrm{orb}}$ and negative $\dot{P}$ measurements, indicating that they are inspiraling~\citep{Israel:2004rb, Strohmayer:2005uc, Strohmayer:2004jk}. If these systems are indeed interacting DWDs undergoing direct impact accretion~\citep{Marsh:2002un, Roelofs:2010uv}, this interpretation is tension with the $\dot{f}_{\mathrm{orb}} > 0$ measurement. This can possibly be explained if they are in the very early stages of mass transfer, when the effect of mass transfer on the angular momentum change is still ``turning on'' and the evolution is still dominated by gravitational radiation. While the evolution model employed here and in previous works~\citep{1988ApJ...332..193V, Marsh:2003rd, 2005astro.ph..8218W, Gokhale:2006yn, Kremer:2017xrg} predicts a short turn-on time of $\sim 100~\mathrm{yr}$ that would make it very unlikely to observe two such systems during this window, more detailed stellar structure calculations coupled to binary evolution predict that this turn-on time can be several orders of magnitude longer~\citep{DAntona:2006wor, Deloye:2007uu, Kaplan:2012hu}, making the AM CVn scenario much more likely for HM Cancri and V407 Vul. Measurements of the acceleration of the orbital frequency of HM Cancri, $\ddot{f}_{\mathrm{orb}}$, indicate that the decrease in the orbital separation is slowing and that the system will turn around and begin outspiraling in $\sim 1200~\mathrm{yr}$~\citep{Strohmayer:2021ien}. These systems are already challenging our understanding of DWD evolution.

The recent success of the Zwicky Transient {Facility} in the discovery of eclipsing DWDs has added significantly to the number of known short-period binaries detectable with LISA and to the number of known AM CVn systems~\citep{Burdge:2020end, 2021AJ....162..113V, 2022MNRAS.509.4171K, 2022MNRAS.512.5440V}. Continued monitoring will reveal whether any of these currently known systems are outspiraling, but a similar method can be used in the future to search archival data taken by the Large Synoptic Survey Telescope~\citep[LSST,][]{2009arXiv0912.0201L} for sources that LISA will be the first to detect~\citep{Korol:2017qcx, Korol:2018wep}. This would allow for the measurement of the frequency derivative for eclipsing sources with $\dot{f}_{\mathrm{GW}}$ outside the limit imposed by the LISA mission duration, expanding the population of DWDs that can be probed in order to learn about the properties of electron-degenerate matter.

\acknowledgements
The authors thank Kevin Burdge, Katie Breivik, Salvatore Vitale, Colm Talbot and Ashley Ruiter for useful discussions. {They also thank the anonymous referee for suggestions that improved the manuscript.} They are also grateful for computational resources provided by the Engaging cluster supported by the Massachusetts Institute of Technology. S.~B. is supported by the National Science Foundation Graduate Research Fellowship under grant No. DGE-1122374. K.K. is supported by an NSF Astronomy and Astrophysics Postdoctoral Fellowship under award AST-2001751. E.T. acknowledges support from the Australian Research Council (ARC) Centre of Excellence CE170100004.

\appendix
\section{Evolution calculation}
\label{ap:evolution}
The angular momentum of a double white dwarf binary changes in time due to three effects: gravitational radiation, tides governing the coupling between the accretor's spin and the orbit, and mass transfer. The total change in angular momentum can thus be expressed as a sum of these three contributions:
\begin{align}
    \dot{J}_{\mathrm{tot}} = \left(-\frac{32G^{3}M_{A}M_{D}(M_{A}+M_{D})}{5c^5a^4} + \sqrt{(1 + q)r_{h}}\frac{\dot{M}_{D}}{M_{D}}\right) J_{\mathrm{orb}} + \frac{kM_{A}R_{A}^{2}\omega}{\tau_{s}}.
    \label{eq:ang_momentum_change}
\end{align}
The first term represents the change in angular momentum due to the emission of gravitational radiation, where $a$ is the orbital separation. The second term represents the contribution from mass transfer following the formulation of \cite{1988ApJ...332..193V}, where the mass ratio $q$ is defined such that $q\equiv M_{D}/M_{A} < 1$, and $r_h$ gives the radius at which the transferred material orbits the accretor for the case of direct-impact accretion in units of the orbital separation, $a$:
\begin{align}
    r_h \approx 0.0883
        + 0.04858\log_{10}(1/q)
        + 0.11489\log_{10}^2(1/q)
        - 0.020475\log_{10}^{3}(1/q).
\end{align}

The third term represents a torque due to the coupling between the accretor's spin and the orbital rotation due to the effect of tides, parameterized in terms of a synchronization timescale, $\tau_{s}$. 
We follow \cite{Marsh:2003rd, 2015ApJ...806...76K} in retaining the scaling for the synchronization time first derived in \cite{1984MNRAS.207..433C}:
\begin{align}
\tau_{s} = \tau_{N}\left(\frac{M_{A}}{M_{D}}\right)^{2}\left(\frac{a}{R_{A}}\right)^{6},
\end{align}
where $\tau_{N}$ is a normalization constant set by the value of the synchronization time at the onset of mass transfer, $\tau_0$,
\begin{align}
    \tau_{N} = \tau_{0}\left(\frac{M_{A,0}}{M_{D,0}}\right)^{-2}\left(\frac{a_0}{R_{A,0}}\right)^{-6}.
\end{align}
We take the initial mass values from our gridded COSMIC population and choose the initial value of the separation such that the donor radius is equal to its Roche lobe radius, meeting the condition for mass transfer to begin. We calculate the radius of each white dwarf using Eggleton's zero-temperature mass radius relation~\citep{1988ApJ...332..193V}:
\begin{align}
R = 0.0114 R_{\odot}
        \left[(M / M_{\mathrm{Ch}})^{-2/3} - (M / M_{\mathrm{Ch}})^{2/3}\right]^{1/2}
         \left[1 + 3.5(M / M_p)^{-2/3} + (M_p / M)\right]^{-2/3},
\end{align}
where the Chandrasekhar mass is given by $M_{\mathrm{Ch}}=1.44~M_{\odot}$ and $M_p = 5.7\times10^{-4}~M_{\odot}$. The Roche lobe radius is given by the approximation of \cite{Eggleton:1983rx}, since we assume the DWDs have no eccentricity throughout the portion of their evolution that we model:
\begin{align}
    R_{L} = r_{L}a, \qquad r_{L} = \frac{0.49 q^{2/ 3}}{0.6q^{2/ 3} + \ln(1 + q ^{1/ 3})},
\end{align}
which allows us to set $a_0 = R_{D,0}/r_{L,0}$.

Because the goal of our analysis is to determine the effect of the strength of tides on the population of mass-transferring DWD systems detectable with LISA, we investigate several different choices for $\tau_{0}$, presenting our main results with $\tau_0=10, 1000,\ \mathrm{and}\ 10^{15}$ years, ranging from strong to weak tides following \cite{Gokhale:2006yn}. The contribution of the tidal term to the rate of change of the angular momentum also scales proportionally to the difference between the accretor's spin frequency and the orbital frequency, $\omega = \Omega_{A} - \Omega_{\mathrm{orb}}$. We assume the donor remains synchronized, and that the initial differential spin is given by
\begin{align}
    \omega_0 = -\tau_0 \dot{\Omega}_{\mathrm{orb},0} =  \frac{3\tau_0\dot{a}}{2}\sqrt{\frac{G(M_{A,0} + M_{D,0})}{a^{5}}},
\end{align}
to simulate the effect of tides prior to the onset of mass transfer following \cite{Marsh:2003rd}. Finally, the moment of inertia constant, $k$, is given by 
\begin{align}
    k = 0.1939(1.44885 - M_{A} / M_{\odot})^{0.1917},
\end{align}
as derived in \cite{Marsh:2003rd} for the Chandrasekhar equation of state for zero-temperature degenerate matter~\citep{1967aits.book.....C}.

The orbital angular momentum is given by
\begin{align}
    J_{\mathrm{orb}} = \left(\frac{Ga}{M_{A}+M_{D}}\right)^{1/2}M_{A}M_{D},
\end{align}
which allows us to rearrange Eq.~\ref{eq:ang_momentum_change} in terms of the orbital separation, 
\begin{align}
    \dot{a} = 2a \left(-\frac{32G^{3}M_{A}M_{D}(M_{A}+M_{D})}{5c^5a^4} - \left[ 1- q - \sqrt{(1 + q)r_{h}}\right]\frac{\dot{M}_{D}}{M_{D}} + \frac{kM_{A}R_{A}^{2}\omega}{\tau_{s}J_{\mathrm{orb}}}\right).
    \label{eq:sep_change}
\end{align}

Next, we turn to the evolution of the accretor's spin. The accretor will spin up due to mass transfer and spin down due to the tidal torque. Stronger tides work to keep the accretor synchronized with the orbit. Thus, the rate of change of the accretor spin is given by~\citep{Marsh:2003rd}
\begin{align}
    \label{eq:accretor_spin}
    \dot{\Omega}_{A} = \dot{\omega} + \dot{\Omega}_{\mathrm{orb}} &= \left(\lambda \Omega_{A} - \frac{GM_{A}r_{h}a}{kR_{A}^{2}}\right)\frac{\dot{M}_{D}}{M_{A}} - \frac{\omega}{\tau_s},\\
    \lambda &= 1 + 2\zeta_A + \frac{d\ln{k}}{d\ln{M}},\\
    \zeta &= \frac{d\ln{R}}{d\ln{M}}.
\end{align}
The accretor can be spun up to its breakup limit, $\Omega_{k} = \sqrt{GM_{A}/R_{A}^{3}}$, when tides are weak and inefficient at resynchronizing the system. In this case, we follow \cite{Marsh:2003rd} in assuming that spin-orbit coupling strengthens once the breakup spin is reached, so that the rate of change of the accretor spin does not exceed the rate of change of the breakup spin, setting
\begin{align}
    \dot{\Omega}_{A} = \dot{\Omega}_k = -\frac{\Omega_{k}\dot{M}_{D}}{2M_{A}}(1-3\zeta_{A}),\ \mathrm{if}\ \dot{\Omega}_{A} > \dot{\Omega}_{k}.
    \label{eq:breakup1}
\end{align}
This corresponds to a modification of the tidal term in Eq.~\ref{eq:accretor_spin}:
\begin{align}
    \frac{\omega}{\tau_{s}'} = \left(\lambda \Omega_{A} - \frac{GM_{A}r_{h}a}{kR_{A}^{2}}\right)\frac{\dot{M}_{D}}{M_{A}} - \dot{\Omega}_{k},
    \label{eq:breakup2}
\end{align}
which we also propagate to the factor of $\omega/\tau_s$ in the last term of Eq.~\ref{eq:sep_change}.

Finally, we discuss the treatment of the accretion rate. We again follow \cite{Marsh:2003rd} and assume an adiabatic accretion rate~\citep{1984ApJ...277..355W},
\begin{align}
    \dot{M}_{D} = 
    \begin{cases}
    -f(M_{A}, M_{D}, a, R_{D})\Delta^{3}, \quad &\Delta > 0\\
    0, \quad &\Delta \leq 0
    \end{cases}
\end{align}
where $\Delta$ is the Roche lobe overfill factor, $\Delta = R_{D} - R_{L}$. The form of $f(M_{A}, M_{D}, a, R_{D})$ is given by~\citep{Marsh:2003rd, 1967aits.book.....C,1977ApJ...211..486W, 1984ApJ...277..355W}
\begin{align}
    f(M_{A}, M_{D}, a, R_{D}) = \frac{8\pi^{3}}{9P}
       \left(\frac{5Gm_e}{h^{2}}\right)^{3 / 2}
        \left(\mu_{e}'m_p\right)^{5 / 2}
        \left(\frac{3\mu M_{A}}{5 r_{L}R_{D}}\right)^{3 / 2}
        [a_D(a_D - 1)]^{-1 / 2},
\end{align}
where $P$ is the orbital period, $m_e$ is the electron mass, $h$ is Planck's constant, $\mu_{e}'=2$ is the mean molecular weight, and
\begin{align}
    \mu = \frac{M_{D}}{M_{A}+M_{D}}, \quad
    a_{D} = \frac{\mu}{x_{L1}^{3}} + \frac{1 - \mu}{(1 - x_{L1})^{3}},\quad
    x_{L1} = 0.5 + 0.22\log_{10}q.
\end{align}
$x_{L1}$ is the distance of the inner Lagrangian point from the center of the donor in units of $a$, taken from \cite{Sepinsky:2007in}. 

As discussed in the main text, our analysis differs from that of \cite{Marsh:2003rd} in the treatment of super-Eddington accretion. We neglect the effect of the asynchronicity of the accretor on the Eddington rate, using the approximation from \cite{1999A&A...349L..17H} without modification:
\begin{align}
\dot{M}_{\mathrm{Edd}} &= \frac{8\pi c m_{p} G M_{A}}{\sigma_T(\phi_{L_1} - \phi_{A})},\\
\phi_{L1} &= -\frac{GM_{D}}{x_{L1}a}
        - \frac{G M_{A}}{a(1 - x_{L1})}
        - \frac{G (M_{A} + M_{D})}{2a^{3}}[a(x_{L1} - \mu/ q)]^{2}\\
\phi_{A} &= -\frac{GM_{D}}{a}
        - \frac{G M_{A}}{R_A}
        - \frac{G (M_{A} + M_{D})}{2a^{3}}\left[\frac{2}{3}R_A^2 + (a- \mu a/q)^{2}\right]
\end{align}
where $m_p$ is the proton mass, $\sigma_T$ is the Thomson cross section, and $\phi_{L_1} - \phi_{A}$ is the difference in the Roche potential at the inner Lagrangian point and the surface of the accretor.

For systems with more extreme mass ratios, which can occur either at birth or via continued mass transfer, the binary separation is large enough that the accreted material will form a disk rather than directly impacting the accretor. This occurs when the stream of accreted material intersects itself at $R_{\min} > R_A$ before reaching the surface of the accretor~\citep{Nelemans:2001nr}:
\begin{align}
    R_{\min} = a(0.04948
        - 0.03815\log q 
        + 0.04752\log^{2} q 
        - 0.006973\log^{3} q).
\end{align}
In the case of disk accretion, we again follow \cite{Marsh:2003rd} and replace $r_h$ in Eqs.~\ref{eq:ang_momentum_change}, \ref{eq:sep_change}, \ref{eq:accretor_spin} with the separation-normalized radius of the accretor, $r_A = R_A/a$. This is a good approximation for disk accretion, where the accreted material comes from the inner edge of the disk near the surface of the accretor.

\section{Oscillations}
\label{ap:oscillations}
For systems near the limit between stable and unstable mass transfer, oscillations can occur in the accretion rate and orbital parameters due to the effect of tides~\citep{Marsh:2003rd, Gokhale:2006yn, 2015ApJ...806...76K}. When mass transfer starts, the accretion rate increases rapidly, leading to a corresponding rapid decrease in the mass transfer timescale. If the mass transfer timescale is shorter than the tidal synchronization timescale, the accretor can be efficiently spun up. The accretion rate continues to increase until tides begin to redistribute angular momentum back to the orbit, which decreases the mass transfer rate and spins down the accretor, and so on. These oscillations are only expected when tidal effects are taken into account, as the orbital period should increase monotonically in time if only mass transfer and GR are included in Eq.~\ref{eq:ang_momentum_change}. See \cite{Gokhale:2006yn} for a more detailed discussion.

For graphical simplicity, we use the \textsc{scipy}~\citep{2020SciPy-NMeth} implementation of a Savitzky-Golay filter~\citep{1964AnaCh..36.1627S} with a window length of 51 samples and a third-order polynomial to smooth the mass transfer and tidal timeseries in Fig.~\ref{fig:evolution} to suppress these oscillations. We only plot the smoothed version to highlight the average behavior of $\dot{f}_{\mathrm{MT}}$ and $\dot{f}_{\mathrm{tides}}$ and use the original timeseries when calculating $\dot{f}_{\mathrm{GW}}$ for our final population in Fig.~\ref{fig:f_fdot}. However, because the oscillations only occur for the first several thousand years after the onset of mass transfer, it is very unlikely that we would catch a binary during this phase of its evolution.

\section{Bayesian inference}
\label{ap:inference}
\subsection{Building the generative model}
\begin{figure*}
\centering
\includegraphics[width=0.45\textwidth]{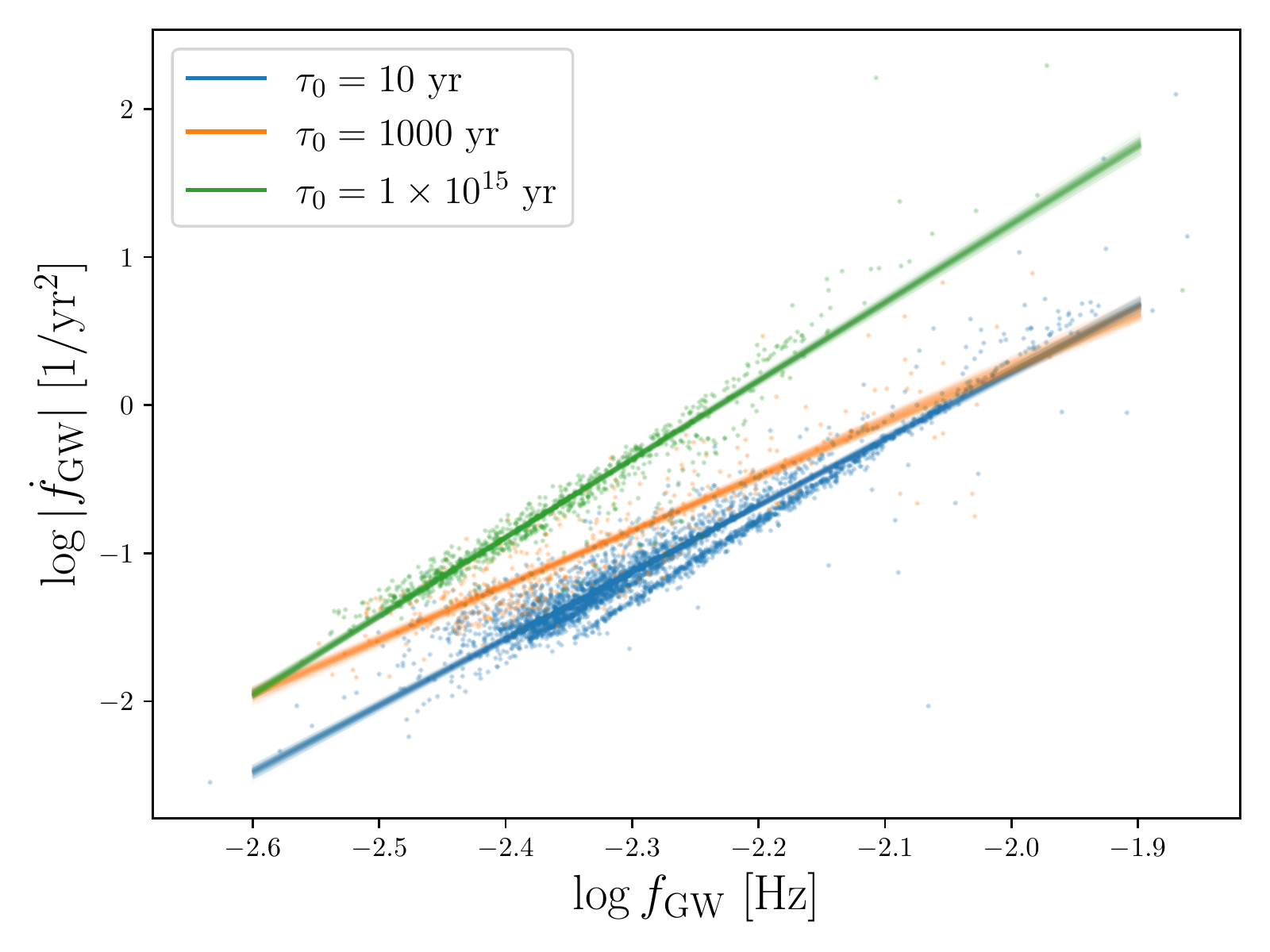}
\includegraphics[width=0.45\textwidth]{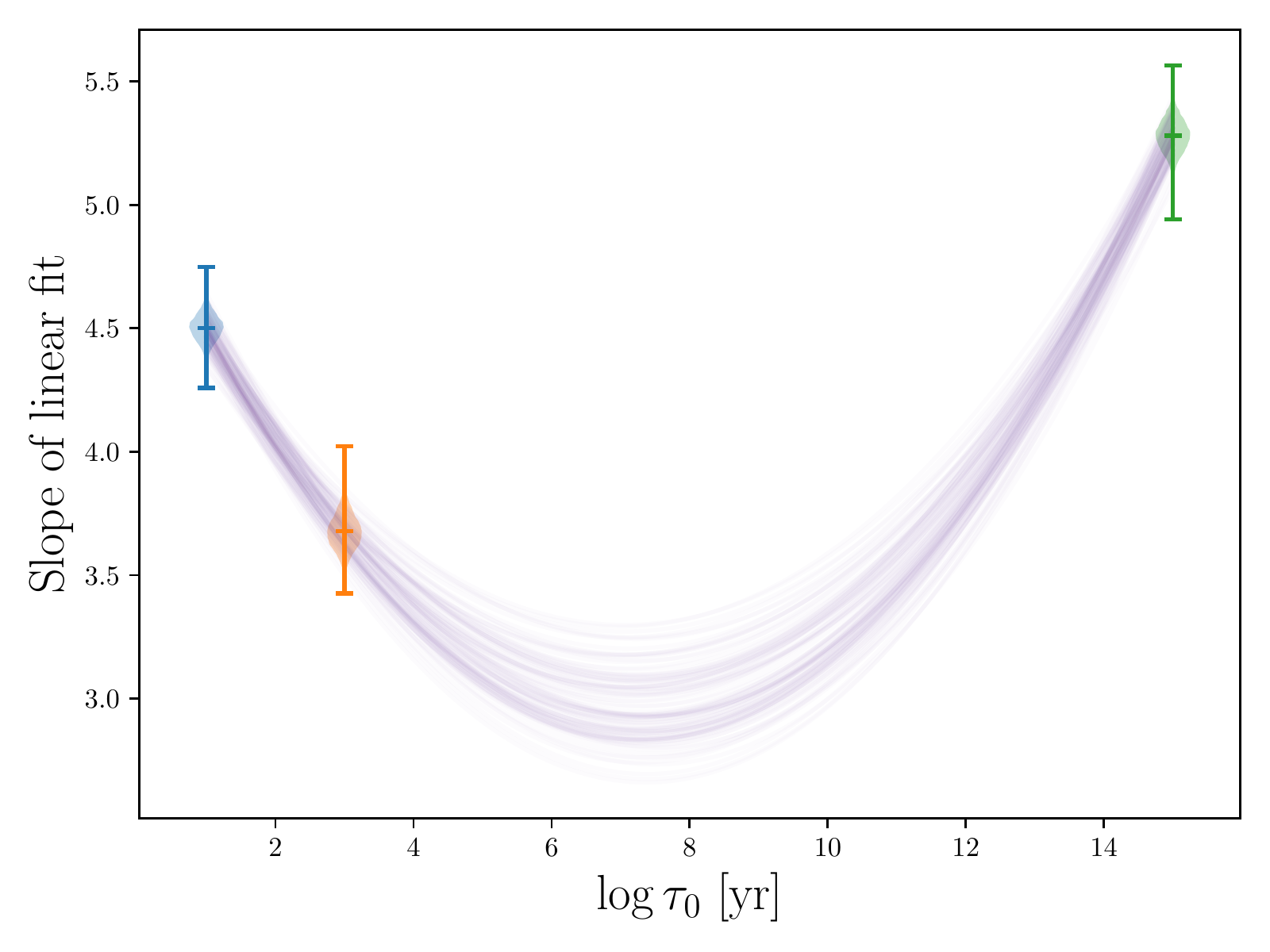}
\caption{Demonstration of the hierarchical model. The left panel shows the $(\log f_{\mathrm{GW}}, \log | \dot{f}_{\mathrm{GW}}|)$ distribution of the training data (scatter plot) with the inferred linear relationship between $\log f_{\mathrm{GW}}$ and $\log | \dot{f}_{\mathrm{GW}}|$ (solid lines) overlaid for each value of $\tau_0$, indicated by the color. The right panel shows the inferred quadratic mapping between the slope of the linear fit and $\log \tau_0$ given in Eq.~\ref{eq:fofa} (purple lines) along with the implied posteriors on $a$ at each value of $\tau_0$ corresponding to the training simulations used to generate the lines in the left panel (violin plots).}
\label{fig:model_demo}
\end{figure*}

{Each of our population synthesis simulations, indexed $j$, generates training data $d_{s,j}$ consisting of individual samples, indexed $k$, in $(\log f_{\mathrm{GW}}, \log \dot{f}_{\mathrm{GW}})$ representing the true values of the frequency and frequency derivative for a particular double white dwarf binary. As described in Section~\ref{sec:inference}, we assume that the correlation between $\log f_{\mathrm{GW}}$ and $\log |\dot{f}_{\mathrm{GW}}|$ can be encapsulated by a linear fit with slope $a$ and intercept $b$.
This allows for a reparametrization of the $(\log f_{\mathrm{GW}}, \log |\dot{f}_{\mathrm{GW}}|)$ distribution in terms of two variables that we assert are independent, $\hat{y}$ and $\hat{z}$:
\begin{align}
    x = \log f_{\mathrm{GW}}, &\quad y = \log |\dot{f}_{\mathrm{GW}}|\\
    \hat{y} = ax + b, &\quad \hat{z} = \mathrm{sign}(x - x_{\min})\sqrt{(x - x_{\min})^{2} + (y - \hat{y}(x_{\min}))^{2}}\\
    \frac{d\hat{z}}{dx}\bigg\rvert_{x=x_{\min}} = 0, &\quad 
    x_{\min} = \frac{x + ay - ba}{1+a^{2}}. \label{eq:xmin}
\end{align}
The variable $\hat{z}$ is defined as the perpendicular distance between the point $(\log f_{\mathrm{GW}}, \log |\dot{f}_{\mathrm{GW}}|) = (x,y)$ and the line defined by $\hat{y}$, and $x_{\min}$ is the $x$-value corresponding to the point on the line with the minimum distance relative to a specific $(x,y)$ sample.
We assume $\hat{z}$ follows a Gaussian distribution with mean $\mu=0$ and unknown width $\sigma$ and that $\hat{y}$ follows a Gamma distribution with unknown shape $k$, scale $\theta$, and location $C$,
\begin{align}
    p(\hat{y}, \hat{z} | \Theta) = \prod_k \mathcal{N}(\hat{z}_{k}: 0, \sigma)\Gamma(\hat{y}_{k}: k, \theta, C),
\end{align}
where $\{a, b, \sigma, k, \theta, C\} \in \Theta$ is the set of parameters describing this latent space. This allows us to write the probability of observing the ensemble $d_{s,j}$ of $(\log f_{\mathrm{GW}}, \log |\dot{f}_{\mathrm{GW}}|)$ for each simulation as
\begin{align}
    \label{eq:likelihood_theta}
    p(d_{s,j} |\Theta) &= \prod_k p(\log f_{\mathrm{GW}, sjk}, \log |\dot{f}_{\mathrm{GW}}|_{sjk} | \Theta)\\
    p(\log f_{\mathrm{GW}, sjk}, \log |\dot{f}_{\mathrm{GW}}|_{sjk} | \Theta) &= p(\hat{y}, \hat{z} | \Theta)\left|\frac{\partial(\hat{z}_{sjk}, \hat{y}_{sjk})}{\partial(\log f_{\mathrm{GW}, sjk}, \log |\dot{f}_{\mathrm{GW}}|_{sjk})}\right|.
    \label{eq:likelihood_theta_individual}
\end{align}
Plugging in the expression for $x_{\min}$ in Eq.~\ref{eq:xmin} and simplifying
\begin{align}
    \hat{z} &= \frac{ax + b - y}{\sqrt{1+a^{2}}}, \quad
    \left|\frac{\partial(\hat{z},\hat{y})}{\partial(x,y)}\right| = \left|\frac{a}{\sqrt{1+a^{2}}}\right|,
\end{align}
we obtain 
\begin{align}
    p(d_{s,j} | \Theta) = \prod_k \mathcal{N}(\hat{z}_{sjk}: 0, \sigma)\Gamma(\hat{y}_{sjk}: k, \theta, C)\left|\frac{a}{\sqrt{1+a^{2}}}\right|.
\end{align}

We now need to determine a mapping between the latent space parameters $\Theta$ physical-space parameter $\tau_0$. We assume that each of the latent space parameters are related to $\log\tau_0$ via either a quadratic or linear function {represented generically as $g(\Theta; \Lambda, \tau_{0})$}
\begin{align}
    \label{eq:fofa}
    g(a; A_a, B_a, C_a, \tau_0) = a &= A_a + B_a\log\tau_0 + C_a\log \tau_0^{2},\\
    b &= A_b + B_b\log\tau_0 + C_b\log \tau_0^{2},\\
    \sigma &= A_{\sigma} + B_{\sigma}\log\tau_0 + C_{\sigma}\log \tau_0^{2},\\
    k &= A_k + B_k\log\tau_0 + C_k\log \tau_0^{2},\\
    \theta &= A_{\theta} + B_{\theta}\log\tau_0, \\
    C &= A_C + B_C\log\tau_0 + C_C\log \tau_0^{2},\\
    \label{eq:fofC}
\end{align}
where we have introduced the hyper-parameters $\Lambda \in \{A_i, B_i, C_i\}\ \forall i \in \Theta$ common to all simulations with different values of $\tau_0$. We want to infer the posteriors for $\Lambda$ given the training data $\{d_{s}\}$ and associated values of initial synchronization timescale for the training simulations, $\{\tau_{s}\}$:
\begin{align}
    \label{eq:lambda_post}
    p(\Lambda | \{d_{s}\}, \{\tau_s\}) &\propto p(\{d_{s}\} | \Lambda, \{\tau_s\})p(\Lambda),\\
    p(\{d_{s}\} | \Lambda, \{\tau_s\}) &=\prod_j p(d_{s, j} | \Lambda, \tau_{s,j}),\\
    p(d_{s, j} | \Lambda, \tau_{s,j}) &= \int p(d_{s, j} | \Theta, \tau_{s,j})p(\Theta | \Lambda, \tau_{s,j})d\Theta,\\
     &= \int p(d_{s, j} | \Theta, \tau_{s,j})\delta(\Theta - g(\Theta; \Lambda, \tau_{s,j}))d\Theta,\\
    &= p(d_{s,j} |g(\Theta; \Lambda, \tau_{s,j})),
    \label{eq:individual_likelihood}
\end{align}
where $g(\Theta; \Lambda, \tau_{s,j})$ is given by the relationships defined in Eqs.~\ref{eq:fofa}-\ref{eq:fofC}. The dependence on $\tau_s$ in $p(d_{s,j} | \Theta, \tau_{s,j})$ enters implicitly via the dependence of $\Theta$ on $\tau_s$, so the final likelihood in Eq.~\ref{eq:individual_likelihood} is given by plugging in the value of $\Theta$ determined for a particular $\Lambda$ draw using $g(\Theta; \Lambda, \tau_{s,j})$ into the probability distribution defined in Eq.~\ref{eq:likelihood_theta}. The term $p(\Lambda)$ in Eq.~\ref{eq:lambda_post} represents the prior on $\Lambda$.

We sample from the posterior in Eq.~\ref{eq:lambda_post} using the \textsc{nestle} sampler and apply uniform priors on all $\Lambda$ parameters with bounds given in Table~\ref{tab:lambda_prior}. We remove the side lobe of samples in $(\log f_{\mathrm{GW}}, \log |\dot{f}_{\mathrm{GW}}|)$ for the $\tau_0=10^{15}~\mathrm{yr}$ simulation from the training data $d_{sjk}$ to preserve the linearity of the relationship between $\log f_{\mathrm{GW}}$ and $\log |\dot{f}_{\mathrm{GW}}|$, as this region is likely artificially overpopulated by our base-population resampling procedure. The posterior on $g(a; \Lambda, \{\tau_{s}\})$ we obtain is shown in the right panel of Fig.~\ref{fig:model_demo}, while the left side shows the training data along with the linear fit inferred for each value of $\tau_{s,j}$.
}

\begin{table}
    \centering
\begin{tabular}{|p{1.5cm} | p{1.5cm} p{1.5cm}|}
    \hline
    Parameter & Minimum & Maximum\\
    \hline
    $A_a$ & 3.76 & 5.76\\
    $B_a$ & -1.001 & 0.001\\
    $C_a$  & 0 & 0.045 \\
    $A_{b}$ & 8.5 & 14.5\\
    $B_{b}$ & -2 & 1\\
    $C_{b}$ & 0.05 & 0.12\\
    $A_{\sigma}$ & 0 & 0.1 \\
    $B_{\sigma}$ & 0.003 & 0.023\\
    $C_{\sigma}$ & -0.003 & 0\\
    $A_{k}$ & -89.86 & 20.14 \\
    $B_{k}$ & 5.28 & 0.72\\
    $C_{k}$ & -0.1 & 0.25\\
    $A_{\theta}$ & 0.08 & 0.32 \\
    $B_{\theta}$ & 0 & 0.016\\
    $A_{C}$ & 0 & 0.1 \\
    $B_{C}$ & 0.003 & 0.023\\
    $C_{C}$ & -0.003 & 0\\
    \hline
\end{tabular}
\caption{Hyper-parameter names and prior bounds. All priors are uniform.}
\label{tab:lambda_prior}
\end{table}

\subsection{Inferring $\tau_0$ for a simulated population}
{We now use the generative model built in the previous section to determine how well $\tau_0$ can be measured using a simulated population with $\tau_0=10^8~\mathrm{yr}$. {Rather than rerunning the computationally-costly mass transfer evolution step for a new value of $\tau_0$, we use the generative model to create simulated data for this value of the tidal synchronization timescale in order to test the mechanics of the model's performance in a part of the parameter space where it was not directly trained. This simulated data will be replaced with real LISA observations in the future (see, e.g., \citealt{2017ApJ...851L..25F, 2017PhRvD..96b3012T} for analogous hierarchical population inference forecasting studies using simulated ground-based gravitational-wave detector data).} Because we do not attempt to build the dependence of the number of stably outspiraling systems observable with LISA on $\tau_0$ into our model, we choose to simulate $N_{\ell}=100$ binaries for demonstration purposes. In order to draw samples from $p(d_{s} | g(\Theta; \Lambda, \tau_0=10^{8}~\mathrm{yr}))$, we need to choose a value of $\Lambda$ to use to calculate $\Theta$. To encapsulate the model uncertainty, we choose a value of $\Lambda$ at random from the posterior obtained in the previous section for each synthetic binary drawn from $p(d_{s} | g(\Theta; \Lambda, \tau_0=10^{8}~\mathrm{yr}))$.

The generative model gives the true values of $(\log f_{\mathrm{GW}}, \log |\dot{f}_{\mathrm{GW}}|)$ for our synthetic population, but we want to incorporate the statistical uncertainty with which these parameters can be measured using LISA observations into our inference of $\tau_0$. To this end, we use the Fisher matrix formalism of \cite{Takahashi:2002ky} to determine the width of the posteriors on $(f_{\mathrm{GW}}, \dot{f}_{\mathrm{GW}})$:
\begin{align}
    \Delta f_{\mathrm{GW}} = 0.22 \left(\frac{10}{\mathrm{SNR}}\right)T_{\mathrm{obs}}^{-1}\\
    \Delta \dot{f}_{\mathrm{GW}} = 0.43 \left(\frac{10}{\mathrm{SNR}}\right)T_{\mathrm{obs}}^{-2}.
\end{align}
This requires us to know the SNR for each binary, which depends on the chirp mass and gravitational-wave frequency. To reduce the computational cost of performing another complete population synthesis simulation with $\tau_0=10^{8}~\mathrm{yr}$ to self-consistently obtain SNRs for our synthetic population, we use \textsc{scipy} to fit the one-dimensional SNR distribution for the simulation with $\lambda=1, \tau_0=10~\mathrm{yr}$ with a $\Gamma$ distribution using the maximum likelihood estimation method and independently sample from this fit to assign SNRs to each of our synthetic events. The inferred $\Gamma$ distribution is also visually consistent with the SNR distributions for the $\tau_0=1000, 10^{15}~\mathrm{yr}$ training simulations. While the SNR observed by LISA is not actually independent of the frequency and frequency derivative of the binary, the simplified approach we take is conservative, since it assigns low SNRs with equal probability for all frequencies. The true distribution including correlations between $f_{\mathrm{GW}}, \dot{f}_{\mathrm{GW}}$ and SNR favors high SNRs for binaries with higher gravitational-wave frequencies, meaning that our approach underestimates the posterior width on $f_{\mathrm{GW}}, \dot{f}_{\mathrm{GW}}$ for some of these high-frequency sources.

With an SNR assigned to each of the 100 sources in our synthetic population, we simulate posteriors on $f_{\mathrm{GW}}, \dot{f}_{\mathrm{GW}}$,
\begin{align}
    \label{eq:gaussian_post}
    p(f_{\mathrm{GW}}, \dot{f}_{\mathrm{GW}} | s) &\propto p(s | f_{\mathrm{GW}}, \dot{f}_{\mathrm{GW}}) p(f_{\mathrm{GW}}, \dot{f}_{\mathrm{GW}})\\
    &\propto \mathcal{N}(f_{\mathrm{GW}}; \mu=f_{\mathrm{GW, obs}}, \sigma=\Delta f_{\mathrm{GW}})\mathcal{N}(\dot{f}_{\mathrm{GW}}; \mu=\dot{f}_{\mathrm{GW, obs}}, \sigma=\Delta \dot{f}_{\mathrm{GW}}),
\end{align}
where $p(s | f_{\mathrm{GW}}, \dot{f}_{\mathrm{GW}})$ is the Gaussian likelihood of observing gravitational-wave strain $s$ given $f_{\mathrm{GW}}, \dot{f}_{\mathrm{GW}}$ and the prior, $p(f_{\mathrm{GW}}, \dot{f}_{\mathrm{GW}})$, is assumed to be uniform. The mean of each posterior is drawn from a Gaussian with the same $\sigma$ peaked at the true values of $f_{\mathrm{GW}}, \dot{f}_{\mathrm{GW}}$ determined by the generative model. This is done to introduce noise into the measurement LISA would obtain for $f_{\mathrm{GW}}, \dot{f}_{\mathrm{GW}}$ for a given system, thereby widening the posterior inferred for $\tau_0$.
In order to match the selection criterion applied to generate our training data, we only include synthetic sources for which this ``observed'' value of the frequency derivative, $\dot{f}_{\mathrm{GW, obs}}$, is constrained to be negative at 95\% credibility, as per its calculated $\Delta \dot{f}_{\mathrm{GW}}$.

In our inference of $\tau_0$ for each synthetic event, we marginalize over both the model uncertainty in the $\Lambda$ posteriors obtained in the previous section and the statistical uncertainty in the measurements of $f_{\mathrm{GW}}, \dot{f}_{\mathrm{GW}}$. We adopt the notation introduced in the previous section and use $d \in \{\log f_{\mathrm{GW}}, \log |\dot{f}_{\mathrm{GW}}|\}$ as shorthand, where the absence of the subscript $s$ indicates that this $d$ consists of the synthetic observations rather than the simulated training data. The posterior for $\tau_0$ for an individual binary with observed gravitational-wave strain $s$ is given by
\begin{align}
    p(\tau_0 | s, \{d_{s}\}, \{\tau_s\}) &\propto p(s | \tau_0, \{d_{s}\}, \{\tau_s\})p(\tau_0),\\
    p(s | \tau_0, \{d_{s}\}, \{\tau_s\}) &= \int p(s | d)p(d | \tau_0, \{d_{s}\}, \{\tau_s\})dd \\
    &= \int p(s | d)p(d | \Theta)p(\Theta | \tau_0, \{d_{s}\}, \{\tau_s\})dd\, d\Theta,\\
   &= \int p(s | d)p(d | \Theta)p(\Theta | \Lambda, \tau_0)p(\Lambda | \tau_0, \{d_{s}\}, \{\tau_s\})dd\, d\Theta\, d\Lambda\\
    &= \int p(s | d)p(d | \Theta)\delta(\Theta - g(\Theta; \Lambda, \tau_0))p(\Lambda | \{d_{s}\}, \{\tau_s\})dd\, d\Theta\, d\Lambda\\
    &= \int p(s | d)p(d | g(\Theta; \Lambda, \tau_0))p(\Lambda | \{d_{s}\}, \{\tau_s\})dd\, d\Lambda.
    \label{eq:tau_likelihood}
\end{align}
The likelihood $p(s|d)$ is related to the Gaussian likelihood in Eq.~\ref{eq:gaussian_post} as
\begin{align}
    p(s | d) = p(s | \log f_{\mathrm{GW}}, \log |\dot{f}_{\mathrm{GW}}|) &= p(s | f_{\mathrm{GW}}, \dot{f}_{\mathrm{GW}})\left|\frac{\partial(\log f_{\mathrm{GW}}, \log |\dot{f}_{\mathrm{GW}}|)}{\partial (f_{\mathrm{GW}}\dot{f}_{\mathrm{GW}})}\right|\\
    &= p(s | f_{\mathrm{GW}}, \dot{f}_{\mathrm{GW}})|f_{\mathrm{GW}}\dot{f}_{\mathrm{GW}}(\ln{10})^{2}|.
\end{align}
We can replace the integrals in Eq.~\ref{eq:tau_likelihood} with sums over samples from $p(s | f_{\mathrm{GW}}, \dot{f}_{\mathrm{GW}})$ in Eq.~\ref{eq:gaussian_post} and from the posterior on the hyper-parameters inferred in the previous section using the training data, $p(\Lambda |\{d_{s}\}, \{\tau_s\})$,
\begin{align}
    p(s | \tau_0, \{d_{s}\}, \{\tau_s\}) = \frac{1}{N_n N_m} \sum_n^{N_n}\sum_m^{N_m} p\left(\log f_{\mathrm{GW},n}, \log |\dot{f}_{\mathrm{GW},n}| \Big| g(\Theta; \Lambda_m, \tau_0)\right)|f_{\mathrm{GW}}\dot{f}_{\mathrm{GW}}(\ln{10})^{2}|,
    \label{eq:tau_likelihood_mc}
\end{align}
where $N_n=1000$ is the number of samples from the posterior on $f_{\mathrm{GW}}, \dot{f}_{\mathrm{GW}}$ and $N_m=5000$ is the number of posterior samples for the hyper-parameters $\Lambda$.

We evaluate the final likelihood in Eq.~\ref{eq:tau_likelihood_mc} on a grid in $\log\tau_0$ and then multiply the individual grids to obtain the posterior on $\tau_0$ for the ensemble of gravitational-wave strain observations corresponding to each of the binaries in our synthetic population, $\{s\}$,
\begin{align}
    p(\tau_0 | \{s\}, \{d_s\}, \{\tau_s\}) = \prod_{\ell}^{N_{\ell}} p(s_{\ell} | \tau_0, \{d_{s}\}, \{\tau_s\}) p(\tau_0),
    \label{eq:tau_post}
\end{align}
assuming a uniform prior on $\log\tau_0$. The posterior on $\log\tau_0$ obtained using Eq.~\ref{eq:tau_post} is shown in Fig.~\ref{fig:tau_post}.
}
\bibliography{refs}
\end{document}